\documentclass[11pt]{article}

\title{Cost-Benefit Analysis of Data Intelligence\\--- Its Broader Interpretations}
\author{
        Min Chen\\
        University of Oxford, UK
}

\date{}

\usepackage[a4paper, total={16cm, 24cm}]{geometry}

\usepackage[pdftex]{graphicx}

\usepackage{mathptmx}
\usepackage{times}

\usepackage{amsfonts}
\usepackage{amsmath}
\usepackage{amssymb}

\DeclareMathAlphabet{\mathcal}{OMS}{cmsy}{m}{n}

\usepackage{enumitem}

\usepackage{url}

\begin{document}
\maketitle

\begin{abstract}
The core of data science is our fundamental understanding about data intelligence processes for transforming data to decisions.
One aspect of this understanding is how to analyze the cost-benefit of data intelligence workflows.
This work is built on the information-theoretic metric proposed by Chen and Golan for this purpose and several recent studies and applications of the metric.
We present a set of extended interpretations of the metric by relating the metric to encryption, compression, model development, perception, cognition, languages, and media.
 
\end{abstract}

\section{Introduction}
\emph{Data intelligence} is an encompassing term for processes such as statistical inference, computational analysis, data visualization, human-computer interaction, machine learning, business intelligence, simulation, prediction, and decision making.
Some of these processes are machine-centric and some others are human-centric, while many are integrated processes that capitalize the relative merits of both.
Recently, Chen and Golan proposed an information-theoretic metric for measuring the cost-benefit of a data intelligence workflow or its individual component processes \cite{Chen:2016:TVCG}.

Several attempts have been made to evidence, falsify, and exemplify this theoretic proposition.
An empirical study was conducted to detect and measure the three quantities of the metric using a visualization process -- a type of human-centric data intelligence process \cite{Kijmongkolchai:2017:CGF}.
The metric was used as the basis for comparing a fully-automated machine learning workflow and a human-assisted machine learning workflow, and for explaining the better models obtained with the latter approach \cite{Tam:2017:TVCG}.
It was applied to the field of virtual environments, relating different dimensions of virtual reality and mixed reality to the cost-benefit metric and offering explanations about the merits and demerits evidenced in practical applications \cite{Chen:2019:VR}.
In addition, a medium-scale elucidation study was conducted to falsify some theories of visualization using some 120 arguments in the literature, and the cost-benefit metric was the only theory that survived the attempts of falsification.

The results of these exercises indicated the explanatory power of this metric in reasoning about the successful and less successful scenarios in several data intelligence workflows, such as visualization, machine learning, and virtual environments, while informing us about some enriched interpretations of this metric in the context of data intelligence as well as some potential interpretations in a broader scope. In this chapter, we report these new insights as an extended elucidation of the cost-benefit metric by Chen and Golan \cite{Chen:2016:TVCG}. We are fully aware that to confirm these interpretations will require a tremendous amount of effort across several different disciplines. Our aim is thus to expound a broader set of interpretations of this particular information-theoretic metric and to stimulate new multi-disciplinary effort in advancing our fundamental understanding about data intelligence.

In the remainder of this chapter, we will first briefly describe the cost-benefit metric and the original interpretations that may be derived from this metric in Section \ref{sec:CBM}.
We will then report two enriched interpretations, in relation to encryption and compression in Section \ref{sec:Cryptography} and to model development in Section \ref{sec:Model}.
We will articulate the potential interpretations, in relation to perception and cognition in Section \ref{sec:Cognition} and to languages and news media in Section \ref{sec:Languages}.
We will offer our concluding remarks in Section \ref{sec:Conclusions}.
 
\section{The Cost-Benefit Metric for Data Intelligence}\label{sec:CBM}
\begin{figure}[h]
\centering
\includegraphics[width=\columnwidth]{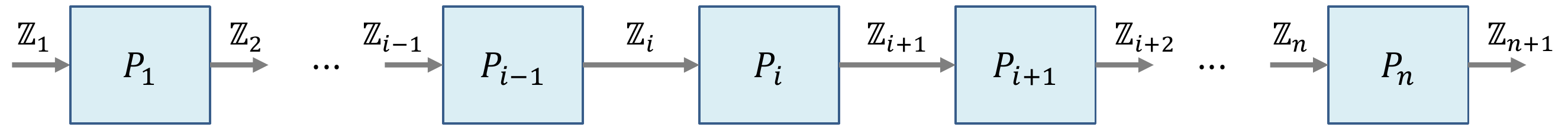}
\caption{A sequentialized representation of a data intelligence workflow.}
\label{fig:DI-basic}
\end{figure}

\paragraph{Processes and Transformations.}
A data intelligence workflow may consist of one or more processes that transform some data to some decisions.
Here the term ``decision'' is a generic placeholder for different types of outcomes that may result from the execution of the workflow, such as identifying an object, an event, or a relation; obtaining a fact, a piece of knowledge, or a collection of views;  selecting a category, place, time, or an option of any arbitrary type; determining a group, a path, or a course of action; or even unconsciously acquiring a memory, an emotion, or a sense of confidence.
The processes in the workflow can be performed by humans, machines, or both jointly.
As shown in Figure \ref{fig:DI-basic}, a sequentialized workflow is, in abstraction, a series of processes $P_1, P_2, \ldots, P_i, \ldots, P_n$.

In theory, the steps in Figure \ref{fig:DI-basic} can be infinitesimally small in time, the resulting changes can be infinitesimally detailed, while the sequence can be innumerably long and the processes can be immeasurably complex.
In practice, one can construct a coarse approximation of a workflow for a specific set of tasks.
Major iterative steps may be sequentialized and represented by temporally-ordered processes for different steps, while minor iterative steps may be combined into a single process.
Parallel processes that are difficult to sequentialize, such as voting by a huge number of people, are typically represented by a single macro process.

As shown in Figure \ref{fig:DI-basic}, all of these processes receive input data from a previous state, process the data, and deliver output data to a new state.
The input data and output data do not have to have a similar semantic definition or be in the same format.
To capture this essence from an information-theoretic perspective, each of these processes in a workflow is referred to as a \emph{transformation}. 

\paragraph{Alphabets and Letters.}
For each transformation $P_i$, all possible input datasets to $P_i$ constitute an input \emph{alphabet} $\mathbb{Z}_i$, and an instance of an input is a  \emph{letter} of this alphabet.
Similarly, all possible output datasets from $P_i$ constitute an output alphabet $\mathbb{Z}_{i+1}$.
In the grand scheme of things, we could consider all possible states of the humans and machines involved in a workflow as parts of these input and output alphabets.
In practice, we may have to restrict the specification of these alphabets based on those types of data that are explicitly defined for each transformation.
Hence many states, such as human knowledge and operating conditions of machines, are usually treated as external variables that have not been encoded in these alphabets. 
The presence or absence of these external variables make the cost-benefit analysis interesting as well as necessary.

\paragraph{Cost-benefit Analysis.}
Consider the transformation from alphabet  $\mathbb{Z}_i$ to $\mathbb{Z}_{i+1}$ at process $P_i$ in Figure \ref{fig:DI-basic}.
Let $\mathcal{H}(\mathbb{Z}_i)$ be the Shannon entropy\footnote{%
Given an alphabet $\mathbb{Z} = \{z_1, z_2, \ldots, z_n\}$ and a probability mass function $P(z)$ for the letters, with the binary logarithm, the Shannon entropy of the alphabet is defined as $\mathcal{H}(\mathbb{Z}) = - \sum_{i=1}^n P(z_i) \log_2 P(z_i)$.}%
~of alphabet $\mathbb{Z}_i$ and $\mathcal{H}(\mathbb{Z}_{i+1})$ be that of $\mathbb{Z}_{i+1}$.
The entropic difference between the two alphabets, $\mathcal{H}(\mathbb{Z}_i) - \mathcal{H}(\mathbb{Z}_{i+1})$, is referred to as \emph{Alphabet Compression}.
Chen and Golan observed a general trend of \emph{Alphabet Compression} in most (if not all) data intelligence workflows, since the decision alphabet is usually much smaller than the original data alphabet in terms of Shannon entropy.

On the other hand, the reduction of Shannon entropy is usually accompanied by the \emph{Potential Distortion} that may be caused by the transformation. Instead of measuring the errors of $\mathbb{Z}_{i+1}$ based on a third-party and likely-subjective metric, we can consider a reconstruction of $\mathbb{Z}_i$ from $\mathbb{Z}_{i+1}$.
If there are external variables (e.g., humans' knowledge) about the data, the context, or the previous transformations, it is possible to have a better reconstruction with the access to such external variables than without.
In \cite{Chen:2016:TVCG}, this is presented as one of the main reasons that explain why visualization is useful.
The potential distortion is measured by the Kullback-Leibore divergence\footnote{%
Given two $n$-letter alphabets $\mathbb{X}$ and $\mathbb{Y}$, and their respective probability mass functions $P(x)$ and $Q(y)$, the Kullback-Leibore divergence of the two alphabets is defined as
$\mathcal{D}_{KL}(\mathbb{X} || \mathbb{Y}) = \sum_{i=1}^n P(x_i) \log_2 P(x_i)/Q(y_i)$.}%
, $\mathcal{D}_{KL}(\mathbb{Z}'_i || \mathbb{Z}_i)$, where $\mathbb{Z}'_i$ is reconstructed from $\mathbb{Z}_{i+1}$.
Furthermore the transformation and reconstruction need to be balanced by the \emph{Cost} involved, which may include the cost of computational and human resources, cognitive load, time required to perform the transformation and reconstruction, the adversary cost due to errors, and so on.
Together the trade-off of these three measures are expressed in Eq.\,(\ref{eq:CBR}): 
\begin{equation}
\label{eq:CBR}
   \frac{\text{Benefit}}{\text{Cost}} = \frac{\text{Alphabet Compression}-\text{Potential Distortion}}{\text{Cost}}
   = \frac{\mathcal{H}(\mathbb{Z}_i) - \mathcal{H}(\mathbb{Z}_{i+1}) - \mathcal{D}_{KL}(\mathbb{Z}'_i || \mathbb{Z}_i) }{\text{Cost}}
\end{equation}

The \emph{Benefit} is measured in the unit of \emph{bit}.
While the most generic cost measure is energy, it can be approximated in practice using a monetary measurement or a time measurement.
This metric suggests several interpretations about data intelligence workflows, all of which can be supported by practical evidence though some may not be instinctively obvious.
These interpretations include:
\begin{enumerate}[label=(\alph*)]
\item
Losing information (i.e., in terms of Shannon entropy) is a ubiquitous phenomenon in data intelligence processes, and has a positive
impact on the \emph{Benefit}.
For instance, numerical regression can be used to transforms an alphabet of $n$ 2D data points in $\mathbb{R}^2$ to an alphabet of $k$ coefficients in $\mathbb{R}$ for a polynomial, and it typically exhibits a positive alphabet compression.
A sorting algorithm transforms an alphabet of $n$ data values to an alphabet of $n$ ordered values, and the former has a higher entropy than the latter.
A user interaction for selecting a radio button out of $k$ choices transforms an alphabet of $\log_2 k$ bits to an alphabet of 0 bits.
\item
 Traditionally losing information has a negative connotation.
This is a one-sided generalization inferred from some observed causal relations that missing useful information leads to difficulties in decision making.
This is in fact a paradoxical observation because the assessment of ``usefulness'' in the cause depends on the assessment of ``difficulties'' in the effect.
The cost-benefit metric in Eq.\,(\ref{eq:CBR}) resolves this self-contradiction by treating \emph{Alphabet Compression} as a positive quality and balancing it with \emph{Potential Distortion} as a negative quality.
\item
When a transformation $P_i$ is a many-to-one mapping from its input alphabet to its output alphabet, the measure of alphabet compression is expected to be positive.
 At the same time, the corresponding reconstruction is a one-to-many mapping, with which potential distortion is also expected.
 Although a typical reconstruction would be based on the maximum entropy principle, this may not be an optimal reconstruction if external variables are present and accessible by the reconstruction (see also (d)).
Assume that the letters in the output alphabet $\mathbb{Z}_{i+1}$ contribute equally towards the subsequent transformations $P_{i+1}, \ldots, P_n$ (see also (f)).
With the same amount of alphabet compression, the more faithful the reconstruction, the better the transformation (see also (e)). 
\item Humans' soft knowledge can be used to reduce the potential distortion.
As discussed earlier, for any human-centric process, if the humans' soft knowledge is not encoded in the input alphabet of the process, such knowledge would be treated as external variables.
For example, consider the process of recognizing a car in an image.
When a large portion of the car is occluded by other objects in the scene, most people can perform this task much better than any automated computer vision algorithm that represents the current state of the art.
This is attributed to the humans' soft knowledge about various visual features of cars and the context suggested by other objects in the scene.
In some decision scenarios, such knowledge may introduce biases into the process.
We will discuss this topic further in Sections \ref{sec:Cryptography}--\ref{sec:Languages}.
\item
Modifying a transformation $P_i$ may change its \emph{Alphabet Compression}, \emph{Potential Distortion}, and \emph{Cost}, and may also change the three
measures in subsequent transformations $P_{i+1}, \ldots, P_n$.
Hence, optimizing the cost-benefit of a data intelligence workflow is fundamentally a global optimization.
However, while it is necessary to maintain a holistic view about the whole workflow, it is both desirable and practical to improve a workflow through controlled localized optimizations in a manner similar to the optimization of manufacturing and businesses processes.
\item
The cost-benefit of a data intelligence workflow is task-dependent. Such tasks are implicitly encoded in the output alphabets of some transformations, usually towards the end of the sequence $P_1, P_2, \ldots, P_n$.
For example, if the task of $P_n$ is to select a final decision from the options defined in its output alphabet $\mathbb{Z}_{n+1}$, $\mathbb{Z}_{n+1}$ thus encodes the essence of the task.
Some potential distortion at an early transformation $P_i\,(i<n)$ may affect this selection but some may not.
\end{enumerate}

\section{Relating the Metric to Encryption and Compression}\label{sec:Cryptography}
Kijmongkolchai et al. reported an empirical study to detect and measure the humans' soft knowledge used in a visualization process \cite{Kijmongkolchai:2017:CGF}.
Their study was designed to evaluate the hypothesis that such knowledge can enhance the cost-benefit ratio of a visualization process by reducing the potential distortion.
It focused on the impact of three classes of soft knowledge: (i) knowledge about application contexts, (ii) knowledge about the patterns to be observed, and (iii) knowledge about statistical measures.
In each trial, eight time-series plots were presented to a participant who was asked to choose a correct time series.
Three criteria were used to define the correctness. They are:
\begin{enumerate}
\item Matching a predefined application context, which may be electrocardiogram (ECG), stock price, and weather temperature.
\item Matching a specific visual pattern, which may be a global pattern of the time series (e.g., ``slowly trending down'', ``wandering base line'', ``anomalous calm'', ``ventricular tachycardia'', etc.) or a local pattern within the time series (e.g., ``sharp rise'', ``January effect'', ``missing a section of data'', ``winter in Alaska'', etc.).
\item Matching a statistical measure, which may be a given minimum, mean, maximum, or standard deviation of the time series. 
\end{enumerate}

Among the eight time-series plots, the seven distractors (i.e., incorrect or partly correct answers) match zero, one, or two criteria.
Before each trial, participants were shown a corresponding newspaper or magazine article where the hints of the application context and specific visual pattern are given.
The article was removed from the trial, for which the participants have to recall the hints featured in the article as soft knowledge.
The corresponding statistical measure was explicitly displayed during the trial as it would be unreasonable to demand such memorization.
The results of the study showed that the participants made 68.3\% correct decisions among all responses by successfully utilizing all three classes of knowledge together.
In comparison with the 12.5\% chance, the positive impact of knowledge in reducing potential distortion was evident.
The work also proposed a mapping from \emph{Accuracy} and \emph{Response Time} collected in the empirical study to \emph{Benefit} and \emph{Cost} in the cost-benefit metric.

During the design of this study, the authors noticed that the eight time series plots presented in each trial do not exhibit the true probability distribution that would in itself lead participants to a correct decision.
Instead, the data alphabet, $\mathbb{Z}_1$, seemed to be intentionally misleading with the maximum entropy of 3 bits (i.e., 12.5\% chance).
Meanwhile, as the corresponding newspaper or magazine article and the statistical measure indicated a correct answer, there was a hidden ``truth distribution''.
This suggests that an extended interpretation is needed to describe the data intelligence workflow in this multiple-choice setup used by numerous empirical studies.

\begin{figure}[t]
\centering
\includegraphics[width=0.9\columnwidth]{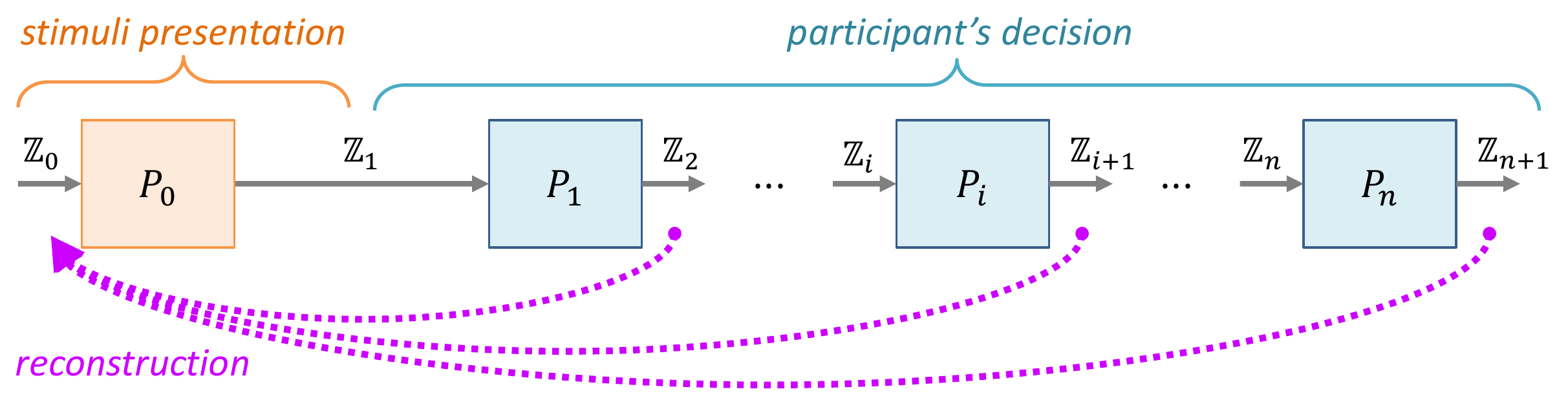}\\
(a) The truth alphabet $\mathbb{Z}_0$ and the pretended alphabet $\mathbb{Z}_1$ in an empirical study.\\[2mm]
\includegraphics[width=0.9\columnwidth]{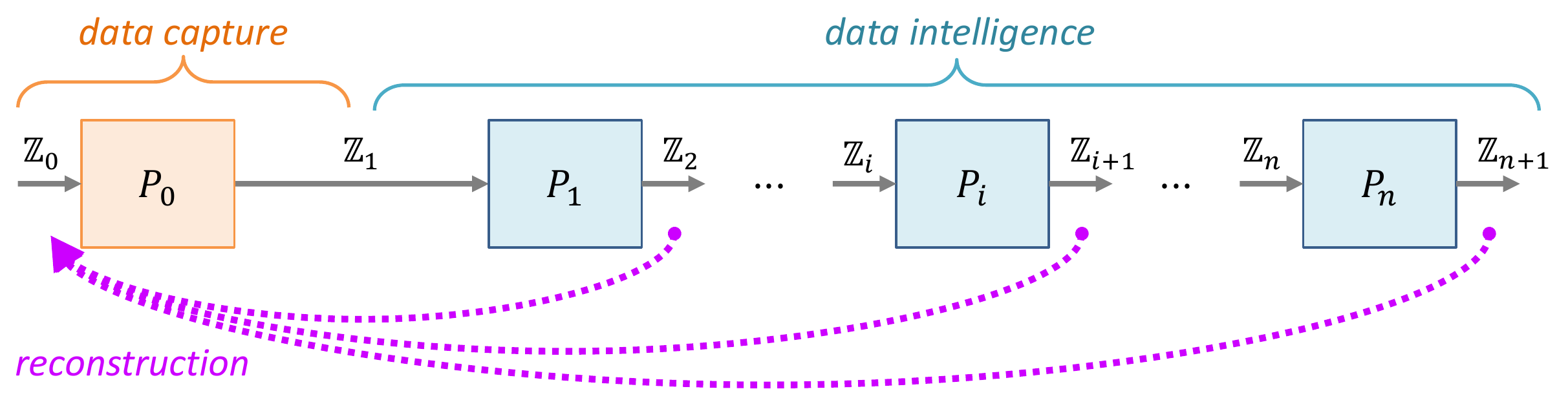}\\
(b) The real-world alphabet $\mathbb{Z}_0$ and the captured data alphabet $\mathbb{Z}_1$ in a data intelligence workflow.
\caption{In an extended interpretation of the cost-benefit metric, a truth alphabet or real-world alphabet $\mathbb{Z}_0$ is introduced, and the quality of the reconstruction ultimately depends on alphabet $\mathbb{Z}_0$ rather than $\mathbb{Z}_1$.}
\label{fig:DI-extended}
\end{figure}
\begin{figure}[t]
\centering
\includegraphics[width=0.9\columnwidth]{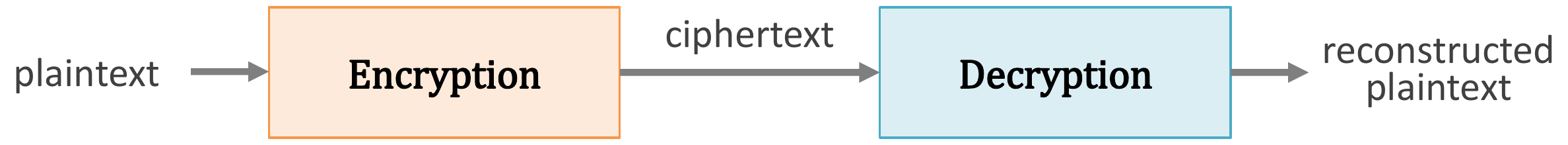}\\
(a) The basic workflow of data encryption and decryption.\\[2mm]
\includegraphics[width=0.9\columnwidth]{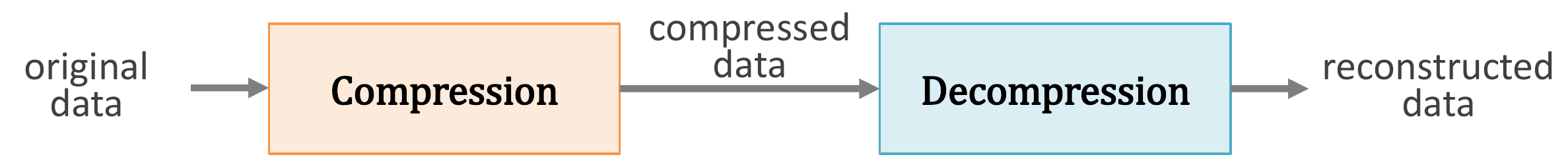}\\
(b) The basic workflow of data compression and decryption.\\[2mm]
\includegraphics[width=0.9\columnwidth]{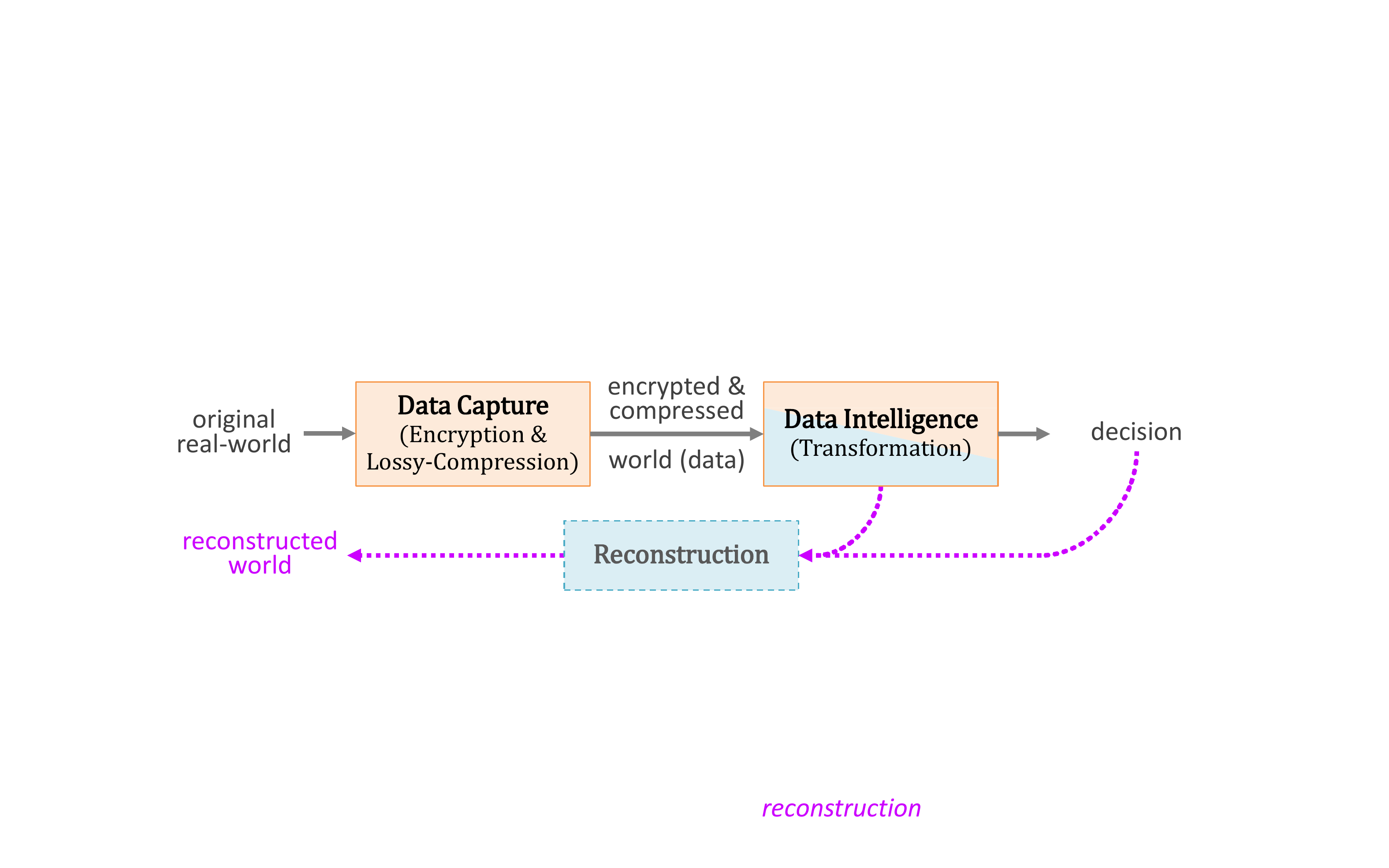}\\
(c) The further abstraction of the data intelligence workflow in Figure \ref{fig:DI-extended}. 
\caption{Juxtaposing the workflows of data encryption, data compression, and data intelligence.}
\label{fig:DI-concept}
\end{figure}


As shown in Figure \ref{fig:DI-extended}(a), it is necessary to introduce a \emph{truth alphabet} $\mathbb{Z}_0$ to encode the hidden truth distribution.
The datasets presented to the participants during the study are from a pretended alphabet $\mathbb{Z}_1$.
The desired reconstruction from the decisions made by the participants should be related to $\mathbb{Z}_0$ but not $\mathbb{Z}_1$.
In other words, the pretended alphabet $\mathbb{Z}_1$ bears some resemblance to an alphabet of ciphertext in cryptography.

This extended interpretation can also be applied to many real-world data intelligence workflows as shown in Figure \ref{fig:DI-extended}(b).
Although alphabet $\mathbb{Z}_1$ in many workflows may not be intentionally misleading, they typically do feature deviation from the truth distribution.
At the same time, in some workflows, such as criminal investigations, alphabet $\mathbb{Z}_1$ likely features some intended distortion of the truth distribution in a manner similar to data encryption.
Hence, the data intelligence workflow should really be about the real-world alphabet $\mathbb{Z}_0$, which the decision alphabet should reflect ultimately.
In addition, the sampled alphabet $\mathbb{Z}_1$ is expected to have much lower entropy than the real-world entropy $\mathbb{Z}_0$.
The data capture process thus exhibits some characteristics of both encryption and lossy-compression.
A data compression method is said to be lossy if the compression process causes information loss and the corresponding decompression process can no longer reconstruct the original data without any distortion.

This naturally suggests that data intelligence is conceptually related to data encryption and data compression, both of which are underpinned by information theory.
Figure \ref{fig:DI-concept} juxtaposes three workflows in their basic forms.
The data capture process transforms a real-world alphabet to an encrypted and compressed world that we refer to as data or sampled data.
The data intelligence process transforms the data alphabet or sampled data alphabet to a decision alphabet, facilitating further compression in terms of Shannon entropy.
Unlike data encryption and data compression, the reconstruction process is not explicitly defined in a data intelligence workflow, possibly because machine-centric processes seldom contain, or are accompanied by, a reverse mapping from an output alphabet to an input alphabet while humans rarely make a conscious effort to reconstruct an input alphabet.
Unconsciously, humans perform this reconstruction all the time as we will discuss later in Section \ref{sec:Cognition}.

The extended interpretation of the cost-benefit metric instigates that the quality of the decisions made by the data intelligence processing block should be measured by the potential distortion in the reconstruction from the decision alphabet to the original real-world alphabet.
Should this reconstruction be defined explicitly, as illustrated in Figure \ref{fig:DI-concept}(c), we would be able to draw a parallel among the three processing blocks for \emph{Encryption}, \emph{Compression}, and \emph{Data Capture}; and another parallel among the three blocks for \emph{Decryption}, \emph{Decompression}, and \emph{Reconstruction}.
Meanwhile, the characterization of the \emph{Data Intelligence} block is multi-faceted.
If this is entirely an automated data intelligence workflow, it exhibits the characteristics of \emph{Compression} as the transformation from a data alphabet to a decision alphabet can be seen as a complex form of lossy-compression. 
If this workflow involves humans who likely perform some forms of reconstruction at some stages, the block also exhibits the characteristics of \emph{Decryption} and \emph{Decompression} in addition.

\section{Relating the Metric to Model Development}\label{sec:Model}
To accompany the proposal of the cost-benefit metric, Chen and Golan also categorized tasks of data analysis and data visualization into four levels \cite{Chen:2016:TVCG} according to the size of the search space for a decision.
The four levels are \emph{Dissemination} of known findings, \emph{Observation} of data, \emph{Analysis} of structures and relations, and \emph{Development} of models.
Using the big O notation in computer science, the search spaces of these four levels are characterized by \emph{constant} O(1), \emph{linear} O($n$), \emph{polynomial} O($n^k$), and \emph{non-deterministic polynomial} (\emph{NP}) (e.g., O($k^n$) or O($n!$)) respectively, where $n$ is the number of letters in the input alphabet $\mathbb{Z}_1$ and $k$ is value such that $1 < k \ll n$.

``Model'' is an overloaded term. Here we use this term strictly for referring to an executable function in the form of $F: \mathbb{Z}_{in} \rightarrow \mathbb{Z}_{out}$.
Hence, a software program is a model, a machine-learned algorithm (e.g., a decision tree and a neural network) is a model, a human's heuristic function is a model, and so on.
When a data intelligence workflow in Figure \ref{fig:DI-basic} is used to obtain a model, the final alphabet $\mathbb{Z}_{n+1}$ consists of all possible models that meet a set of conditions in a specific application context.
Although the number of letters in a typical model alphabet seems to be extraordinarily large, many letters do not meet the predefined conditions in the specific application context.
Hence, the actual entropy of $\mathbb{Z}_{n+1}$ would be much lower than the maximal amount of entropy derived by enumerating all combinations of the model components.
For example, the number of all possible programs with $n$ or fewer lines of code, the number of all possible neural networks with $m$ or fewer neurons, or the number of all possible Bayesian networks with $l$ or fewer nodes and probability values rounded off to $k$ decimal digits would be intractable, but there will only be proportionally a much smaller subset of programs, neural networks, or Bayesian networks that meet the predefined requirements for such a model in a specific application context.
However, the main challenge in model development is that the strategy for finding an acceptable model while filtering out the extraordinarily large number of unacceptable models is often not well-defined or ineffective.
For example, one often refers to programming as ``art'' and parameter tuning as ``black art'', metaphorically reflecting the lack of well-defined or effective strategy in dealing with the NP search space.

Consider an alphabet $\mathbb{M}_{all}$ that contains all possible functions that we can create or use.
Whether we like it or not, some of these functions are being created using \emph{machine learning} (\emph{ML}).
This is not because ML can develop better algorithms or more reliable systems than trained computer scientists and software engineers.
ML is merely a model-developmental tool that helps us write an approximate software function, for which we do not quite know the exact algorithm or would take an unaffordable amount of time to figure it out.
In the systems deployed in practical environments, only a small number of components are machine-learned algorithms.
Nevertheless, this model-developmental tool is becoming more and more powerful and useful because of the increasing availability of training data and high performance computing for optimization.
With such a tool, we can explore new areas in the \emph{space of functions}, $\mathbb{M}_{ML} \subset \mathbb{M}_{all}$, which is programmable on a stored program computer and where conventional algorithms are not yet found or effective.

\begin{figure}[t]
\centering
\includegraphics[width=\columnwidth]{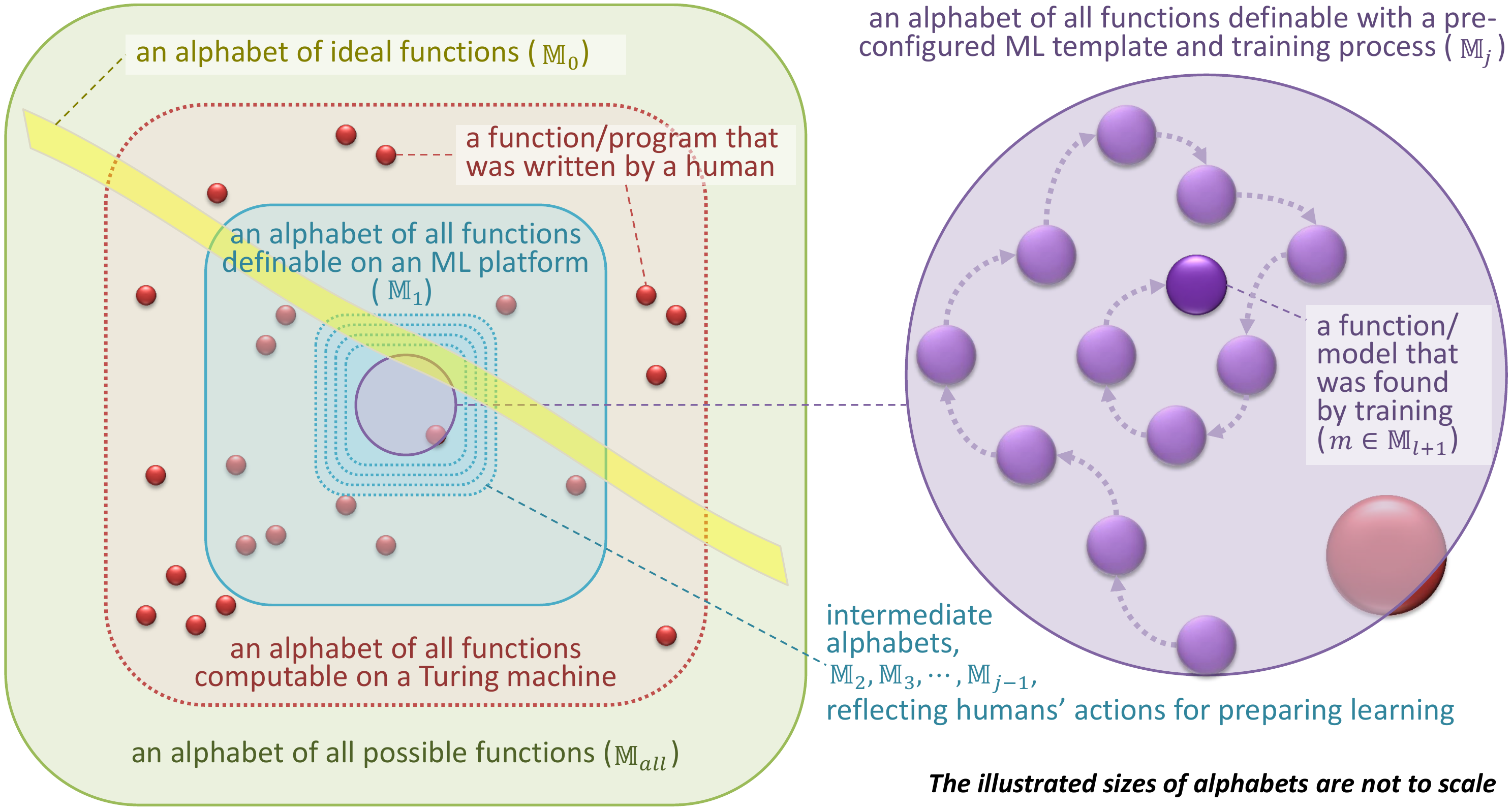}
\caption{A machine learning (ML) workflow searches for a model in the alphabet $\mathbb{M}_1$ determined by an underlying platform. Human- and machine-centric processes work together to enable the search by gradually reducing the size of the search space, which can be measured  by the entropy of the corresponding alphabet. The searching is not assured to find an ideal model in $\mathbb{M}_0$, but can examine and test automatically numerous candidature models in a small space. The small search space $\mathbb{M}_j$ is pre-defined by humans, while the search path is determined by the training data and some control parameters.}
\label{fig:ML-space}
\end{figure}

In theoretical computer science, a programming language is said to be \emph{Turing complete} if it can describe any single-taped Turing machine and vice versa.
We can consider an underlying platform (or framework) used by ML (e.g., neural networks, decision trees, Bayesian networks, support vector machines, genetic algorithms, etc.) as a programming language.
In such a language, there is a very limited set of constructs (e.g., node, edge, weight).
The space of $\mathbb{M}_{ML}$ is a union of all underlying platforms.
A ML workflow typically deals with only one such platform or a very small subset of platforms, which determines an alphabet $\mathbb{M}_1$ (Figure \ref{fig:ML-space}), i.e., the space of all functions definable on the chosen platform(s).

For any practical ML applications, humans usually do the clever part of the programming by defining a \emph{template}, such as determining the candidature variables for a decision tree, the order of nodes in a neural network, or the possible connectivity of a Bayesian network.
The ML tool then does the tedious and repetitive part of the programming for fine-tuning the template using training data to make a function.
It is known that most commonly used ML platforms, such as forward neural networks, decision trees, random forests, and static Bayesian networks are not Turing complete.
Some others are Turing complete (e.g., recurrent neural networks and dynamic Bayesian networks), but their demand for training data is usually exorbitant.
For any ML model, once a template is determined by the humans, the \emph{parameter space} that the ML tool can explore is certainly not Turing complete.
This implies that the ``intelligence'' or ``creativity'' of the ML tool is restricted to the tedious and repetitive part of the programming for fine-tuning the template.

\begin{figure}[t]
\centering
\includegraphics[width=\columnwidth]{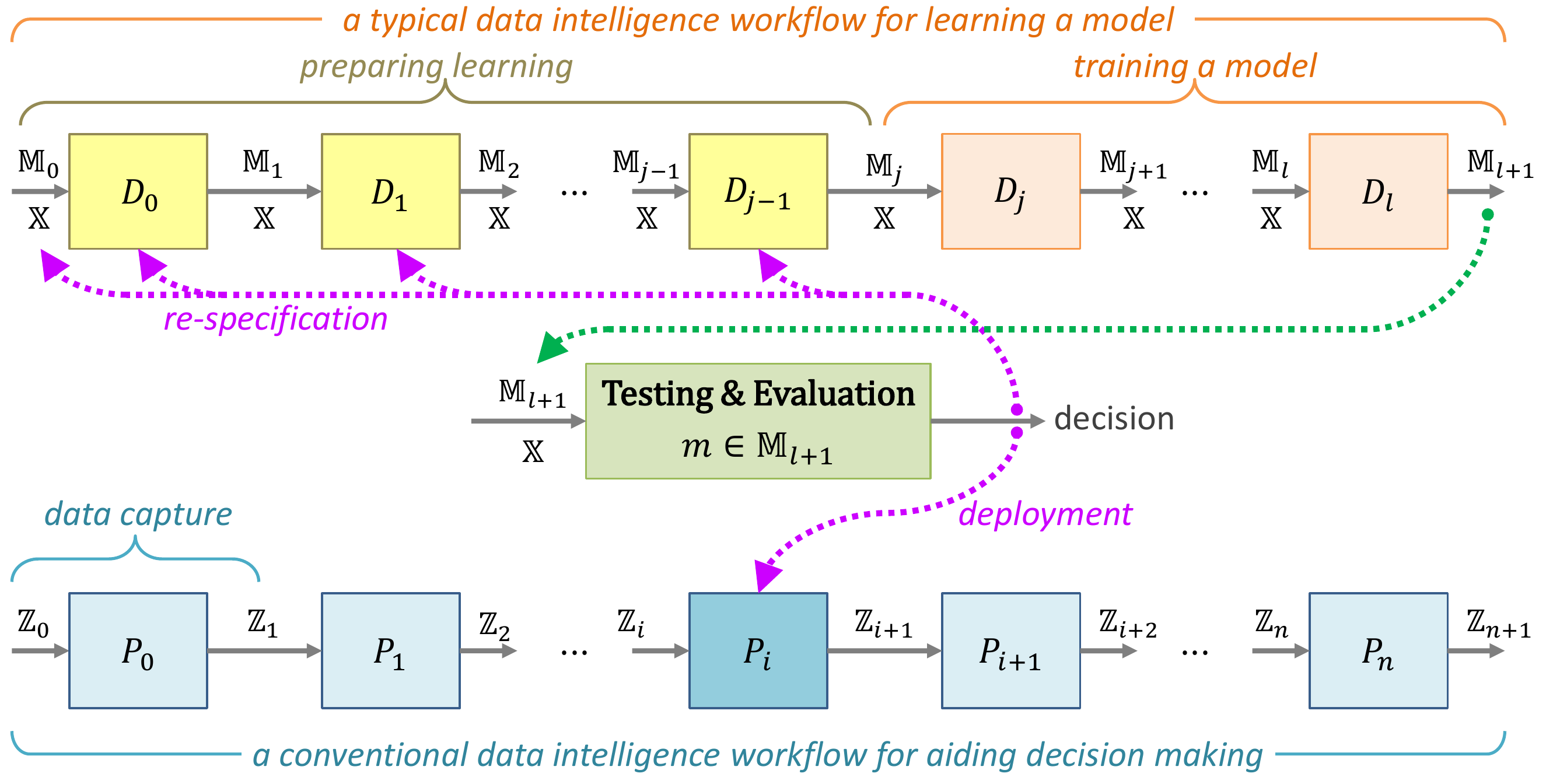}
\caption{The relationship between a typical machine learning workflow (e.g., for learning a decision tree, a neural network, a Markov chain model, etc.) and a conventional data intelligence workflow (e.g., for aiding anomaly detection, market analysis, document analysis, etc.).}
\label{fig:DI-ML}
\end{figure}

Figure \ref{fig:DI-ML} illustrates a typical workflow for supervised learning, and its relationship with a conventional data intelligence workflow where a learned model will be deployed as an intermediate step $P_i$.
In the conventional data intelligence workflow, each of the other processes, i.e., $P_w (w\neq i)$, can be either a machine- or human-centric process.
For example, $P_0$ may be a process for capturing data from a real world environment (e.g., taking photographs, measuring children's heights, conducing a market survey, and so on).
$P_1, \ldots, P_{i-1}$ may be a set of processes for initial observations, data cleaning, statistical analysis, and feature computation.
$P_i$ may be an automated process for categorizing the captured data into different classes.
$P_{i+1}, \ldots, P_n$ may be a set of processes for combining the classified data with other data (e.g., historical notes,  free-text remarks, etc.), discussing options of the decision to be made, and finalizing a decision by voting. 
 
Let $\mathbb{X}$ be an alphabet of annotated data for training and testing.
It is commonly stored as pairs of datasets and labels corresponding to the input and output alphabets of $P_i$ respectively.
In other words, each $x \in \mathbb{X}$ is defined as a pair $(\alpha, \beta)$ such that $\alpha \in \mathbb{Z}_i$ and $\beta \in \mathbb{Z}_{i+1}$.
It is helpful to note that we do not prescribe that the datasets used for ML must be the raw data captured from the real world, i.e., $\mathbb{Z}_1$.
This is because (i) a machine-learned model is rarely the sole function in a data intelligence workflow and there may be other transformations before $P_i$ for pre-processing, analyzing, and visualizing data; (ii) captured datasets are often required cleaning before they can be used as part of the data alphabet $\mathbb{X}$ for training and testing and as the input alphabet $\mathbb{Z}_i$ to the machine-learned model $P_i$; and (iii) it is common to extract features from captured datasets using manually-constructed programs and to use such feature data in $\mathbb{X}$ and $\mathbb{Z}_i$ instead of (or in addition to) the captured datasets. 

Unlike a conventional data intelligence workflow that transforms the data alphabets $\mathbb{Z}_i$ (as shown in the lower part of Figure \ref{fig:DI-ML}), the ML workflow (shown in the upper part) transforms the model alphabets $\mathbb{M}_j, j=0, 1, \ldots, l+1$, while making use of different letters in the same data alphabet $\mathbb{X}$.
The ML workflow consists of two major stages, \emph{preparing learning} and \emph{model learning}.
The stage of preparing learning, which is the clever programming part and is illustrated in Figure \ref{fig:DI-ML} as $D_0, D_1, \ldots, D_{j-1}$, consists primarily of human-centric processes for selecting an underlying platform, constructing a template, specifying initial conditions of the template, and setting platform-specific control parameters that influence the performance of the processes in the model learning stage.

The stage of model learning is the tedious and repetitive part of the workflow and is illustrated as $D_j, \ldots, D_l$ in Figure \ref{fig:DI-ML}. 
It consists primarily of machine-centric processes that are pre-programmed to construct a model (or a set of models) step by step with some platforms (e.g., decision trees, random forest, and Markov chain models) or refining model parameters iteratively with other platforms (e.g., regression, neural networks, and Bayesian networks).
The approach of parameter refinement is illustrated on the right of Figure \ref{fig:ML-space}, where the model learning stage starts with an initial set of parameters (e.g., the coefficients of a regression model, connectivity of a neural network, probability values in a  Bayesian network).
The processes, $D_j, D_{j+1}, \ldots, D_l$, at this stage modify these model parameters according to the training datasets in $\mathbb{X}$ encountered as well as the pre-defined platform-specific control parameters.
As a model is defined by a template and a set of model parameters, the progressive changes actualized by these processes steer the search for an optimal model (or a set of optimal models) in the alphabet $\mathbb{M}_j$.
This mostly-automated stage is often mistaken or misrepresented as the entire data intelligence workflow for machine learning, largely because the more credits are given to the machine-centric processes, the better artificial intelligence appears to be.

After a model (or a set of models) is learned,  it is passed onto an independent process for testing and evaluation under the supervision of humans as shown in the center of Figure \ref{fig:DI-ML},
When a learned model is considered to be unsatisfactory, the humans' effort is redirected to the processes in the stage of preparing learning, where a template may be modified and control parameters be adjusted.
In some cases, the training and testing alphabet $\mathbb{X}$ is considered to be problematic, and this may lead to further activities in data capture, data cleaning, feature specification, algorithm designs for feature extraction, and so on.
The processes for modifying $\mathbb{X}$ are normally a subset of the processes proceeding $P_i$, that is, $P_0, P_1, \ldots, P_{i-1}$ in the conventional data intelligence process.

Information-theoretically, we may follow the discussions in Section \ref{sec:Cryptography} by considering the initial alphabet $\mathbb{M}_0$ as a collection of ideal models that can suitably be deployed as $P_i$ for transforming $\mathbb{Z}_i$ to $\mathbb{Z}_{i+1}$.
We can consider the first process $D_0$ in the data intelligence workflow for learning a model as an encryption by selecting a specific platform, since $\mathbb{M}_1$ consists of all models that can be constructed under this platform, it not only conceals the truth alphabet $\mathbb{M}_0$ but also seldom leads to the discovery of an ideal model in $\mathbb{M}_0$.
Meanwhile, selecting a platform facilitates a significant amount of alphabet compression from the space of all possible functions $\mathbb{M}_{all}$.
As illustrated in Figure \ref{fig:ML-space}, when such a compression is lossy, $\mathbb{M}_1$ cannot guarantee that any ideal model in $\mathbb{M}_0$ is recoverable.
Further alphabet compression, in a massive amount of entropy, is delivered by the human-centric processes $D_1, \ldots, D_{j-1}$.
The candidature models that are contained in $\mathbb{M}_j$ are substantially constrained by the particular template and the control parameters that determine the search strategy and effort.

During the model learning stage, the training data alphabet $\mathbb{X}$ is used to reduce the entropy of $D_{j-1}$ while minimizing the potential distortion.
With construction-based platforms, alphabet compression from $\mathbb{M}_j$ to $\mathbb{M}_{l+1}$ is achieved gradually in each construction step (e.g., determining a node in a decision tree or a Markov chain model).
With parameter-refinement platforms, the entropy of interim alphabets $\mathbb{M}_{j+1}, \ldots, \mathbb{M}_l$ may reduce gradually if the training manages to converge, but may remain more or less at the same level as $\mathbb{M}_j$ if it fails to converge.
Whatever the case, the alphabet $\mathbb{M}_{l+1}$ resulting from the model learning stage typically contains only one or a few models.

In general, the data intelligence workflow for learning a model exhibits the same trend of alphabet compression.
The stage of preparing learning consists of knowledge-driven processes for reducing the Shannon entropy of the model alphabet, while human knowledge and heuristics are utilized to minimize the potential distortion.
The stage of model learning is largely data-driven, where the training algorithm pursues the goal of alphabet compression while various metrics (e.g., the impurity metric in decision tree construction or fitness function in genetic algorithms) are used to minimize the potential distortion.
While the automation in the stage of model learning is highly desirable for cost reduction, it appears to have difficulties to replace the human-centric processes in the stage of preparing learning.
Hence a trade-off has been commonly made between the costs of human resources (e.g., constructing a template in $D_1$) and the computational cost for searching in a significantly larger alphabet (e.g., $\mathbb{M}_1$).
The independent testing and evaluation process, together with the feedback loops to $\mathbb{X}, D_0, D_1, \ldots, D_{j-1}$, provides ways to reduce potential distortion while incurring further cost.

Many in the field of ML also encounter the difficulties in obtaining a suitable training and testing alphabet $\mathbb{X}$ that would be a representative sample of the relations between $\mathbb{Z}_i$ and $\mathbb{Z}_{i+1}$.
In practice, alphabet $\mathbb{X}$ is often too sparse or has a skewed probability distribution.
Some in ML thus introduce human knowledge into the stage of model learning to alleviate the problem.
Tam et al. \cite{Tam:2017:TVCG} investigated two practical case studies where human knowledge helped derive better models than fully-automated processes at the stage of model learning.
They made the computational processes as the ``observers'', and the humans' decisions as the information received by the ``observers''. By measuring the amount of uncertainty removed due to such information, they estimated the amount of human knowledge available to the stage of model learning.

They considered human knowledges in two categories, namely, \emph{soft alphabets} and \emph{soft models}.
Soft alphabets are referred to variables that have not been captured in the data, for example, in their case studies, the knowledge about which facial feature is more indicative about a type of emotion or which feature extraction algorithms are more reliable.
Soft models are referred to human heuristic processes for making some decisions, for which the ML workflow does not yet have an effective machine-centric process. These soft models do not have a pre-defined answers but respond dynamically to inputs as a function $F: \mathbb{Z}_{in} \rightarrow \mathbb{Z}_{out}$.
For example, in their case studies, (i) given a facial photo (input), a human imagines how the person would smile (output); (ii) given a video featuring a facial expression (input), a human judges if the expression is genuine or unnatural (output); (ii) given a video (input), a human determines if it is an outlier or not (output); (iii) given a set of points on an axis (input), a human decides how to divide the axis into two or a few sections based on the grouping patterns of the points (output); and (iv) given a section of an axis with data points of different classes, a human predicts if the entanglement can be resolved using another unused axis.
They discovered that in all of their case studies, the amount of human knowledge (measured in bits) available to the stage of model learning is much more than the information contained in the training data (also measured in bits).
Their investigation confirmed that human knowledge can be used to improve the cost-benefit of the part of the ML workflow that is traditionally very much machine-centric.

Recently, Sacha et al. developed an ontology that mapped out all major processes in ML \cite{Sacha:2019:TVCG}.
Built on a detailed review of the relevant literature, this ontology confirms that there are a large number of human-centric processes in ML workflows.
While recognizing humans' role in ML does not in any way undermine the necessity for advancing ML as a technology, scientifically such recognition can lead to better understanding about what soft alphabets are not in the training data and what soft models are not available to the automated processes.
Practically, it can also stimulate the development of new technical tools for enabling humans to impart their knowledge in ML workflows more effectively and efficiently.

Chen et al.  employed the cost-benefit metric to analyze a wide of range of applications of virtual reality and virtual environments, such as theatre-based education systems, real-time mixed reality systems, ``big data'' visualization systems, and virtual reality training systems.
In particular, they examined applications in medicine and sports, where humans' soft models are being learned through virtual reality training systems \cite{Chen:2019:VR}.
In these applications, the primary reason for using virtual reality is the lack of access to the required reality $\mathbb{R}$.
For example, it would be inappropriate to train certain medical procedures on real patients.
So the training and testing alphabets $\mathbb{X}$ is simulated using virtual reality.
While the performance of the ideal models in $\mathbb{M}_0$ can be anticipated and described easily, we have very limited knowledge about the configurations and mechanisms of the models under training (i.e., $m \in \mathbb{M}_1, \mathbb{M}_2, \ldots, \mathbb{M}_j, \ldots,  \mathbb{M}_l,  \mathbb{M}_{l+1}$).
One may prepare a participant in the training with many instructions at the first stage, and the participant may train against $\mathbb{X}$ in numerous iterations in the second stage.
However, with limited understanding of the inner-workings of the models, the cost-benefit of this form of model development depends hugely on the trade-off of the potential distortion in relating virtual reality $\mathbb{X}$ back to the reality $\mathbb{R}$ and the costs for producing $\mathbb{X}.$

Chen et al. used information-theoretic analysis to confirms the cost-benefit of this form of human model development \cite{Chen:2019:VR} .
They drew from evidence in cognitive science to articulate that human-centric models for motor coordination are more complex than most, if not all, current machine-centric models.
They drew evidence from practical applications to articulate the usefulness and effectiveness of training human models using virtual reality.
They pointed out the need to use conventional data intelligence workflow to help understand the inner-workings of models under training, and to use such understanding to optimize the design of $\mathbb{X}$.

\section{Relating the Metric to Perception and Cognition}\label{sec:Cognition}
The human sensory and cognitive system is an intelligent system that processes a huge amount of data every second.
For example, it is estimated that our two eyes have some 260 million of photoreceptor cells that receive lights as input data.
If the light that arrives at each cell is a variable, there are some 260 million input variables.
Most televisions and computer displays were designed with an assumption that human eyes can receive up to 50-90 different images per second, that is, about 11-20 milliseconds per image.
Some perception experiments showed that human eyes can recognize an image after seeing it for 13 milliseconds.
So within a second, the human visual system can potentially process more than 13 billion pieces of data ($13 \times 10^9 = 50 \textit{ images per second} \times 260 \times 10^6\textit{ input variables}$).

Cognitive scientists have already discovered a number of mechanisms that enable the human visual system to perform alphabet compression almost effortlessly.
One of such mechanisms is \textbf{selective attention}, which enables humans to distribute limited cognitive resources to different visual signals non-uniformly according to their anticipated importance.
In most situations, it works cost-beneficially.
For example, when we drive, we do not pay equal attention to all visual signals appearing in various windows and mirrors.
We focus mostly on the visual signals in the front windscreen with some peripheral attention to those signals in the front side windows as well as the rear-view and wing mirrors.
For those observed signals, we concentrate on those related to road conditions and potential hazards.
At any moment, a huge amount of visual signals available to our eyes do not receive much attention.
There is thus a non-trivial amount of potential distortion in reconstructing the actual scene.
We are unlikely to notice the color and style of every pedestrian's clothing.
Such omission is referred to as \emph{inattentional blindness}.
Occasionally such omission may inattentionally miss out some critical signals, causing an accident.
However, on balance, selective attention is cost-beneficial as it enables us to keep the cognitive load (i.e., cost) low and to maintain sufficient processing capability for the new visual signals arriving continuously.

\begin{figure}[h]
\centering
\includegraphics[width=\columnwidth]{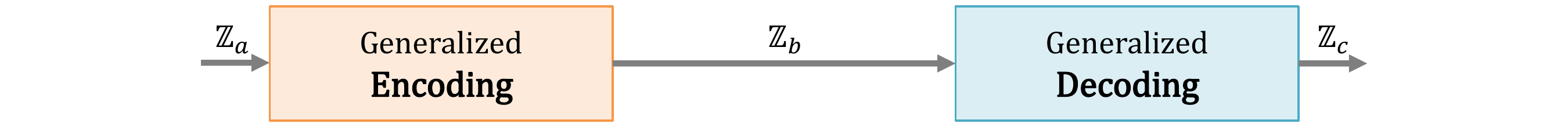}\\
(a) a generic data intelligence workflow in human mind\\[2mm]
\includegraphics[width=\columnwidth]{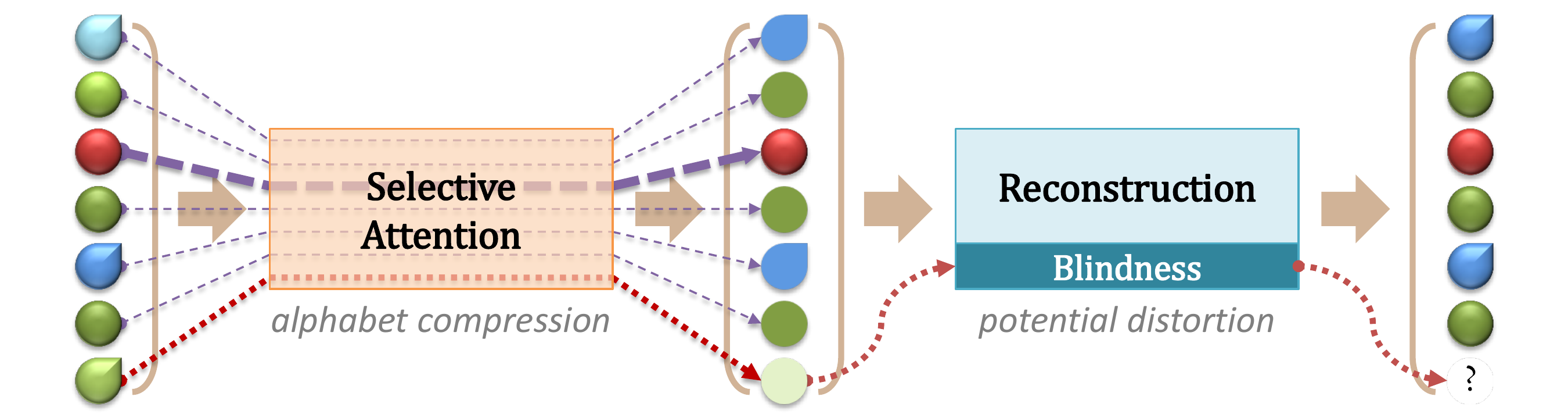}\\
(b) selective attention vs. inattentional blindness\\[2mm]
\includegraphics[width=\columnwidth]{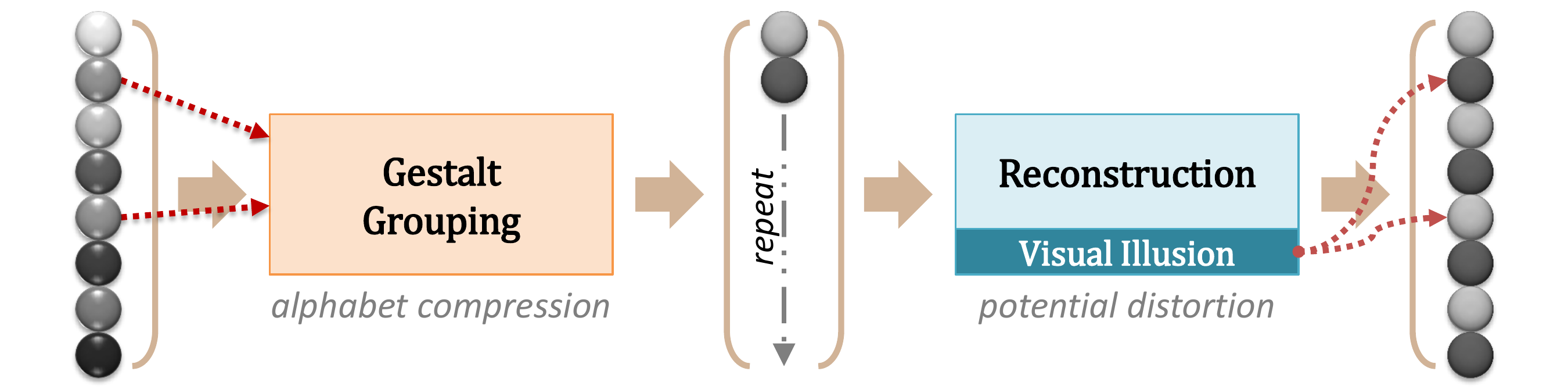}\\
(c) gestalt grouping vs. visual illusion
\caption{\emph{Selective attention} and \emph{gestalt grouping} are two major phenomena in human perception. Information-theoretically, they feature significant alphabet compression and are cost-beneficial, but may occasionally cause potential distortion in the forms of \emph{inattentional blindness} and \emph{visual illusion}.}
\label{fig:DI-Psy1}
\end{figure}

Figure \ref{fig:DI-Psy1}(a) shows a schematic representation of a family of perceptual and cognitive workflows, where we generalize the terms \emph{encoding} and \emph{decoding} from their narrow interpretation of converting the representations of data from one form to another.
Here, ``Encoding'' is a transformation from an input alphabet $\mathbb{Z}_a$ to an output alphabet $\mathbb{Z}_b$, where $\mathbb{Z}_b$ can be an alternative representation of $\mathbb{Z}_a$ as well as a derived alphabet with very different semantics.
It includes not only the traditional interpretations such as encoding, compression, and encryption, but also many broad interpretations such as feature extraction, statistical inference, data visualization, model developments, and some data intelligence processes in the human mind as we are discussing in this section.
On the other hand, ``Decoding'' is an inverse transformation, explicitly or implicitly defined, for an attempt to reconstruct alphabet $\mathbb{Z}_a$ from $\mathbb{Z}_b$.
As the prefect reconstruction is not guaranteed, the decoding results in $\mathbb{Z}'_a$ that has the same letters as $\mathbb{Z}_a$ but a different probability distribution. 

As illustrated in Figure \ref{fig:DI-Psy1}(b), \emph{selective attention} is an instance of encoding.
When a person focus his/her attention on the red object, for example, the process allows more signals (e.g., details of geometric and textual features) to be forwarded to the subsequent cognitive processes.
Meanwhile, other objects in the scene receive less attention, and the process forwards less or none of their signals.  
Selective attention clearly exhibits alphabet compression, and inevitably may cause potential distortion in an inverse transformation.
For example, should the person close his/her eyes and try to imagine the scene of objects, some objects may be incorrectly reconstructed or not reconstructed at all.
Often the person would use his/her knowledge and previous experience about the scene and various objects to fill in some missing signals during the reconstruction.
\emph{Inattentional blindness} actually refers to a scenario where a specific object in the scene or some features of the object that ``should have been'' correctly reconstructed happens to be incorrectly reconstructed or not reconstructed at all.
As there are numerously objects and features are not correctly reconstructed, the criteria for ``should have been'' depend on the objectives of the attentional effort.
In many cognitive experiments for evidencing inattentional blindness, participants' attentional effort is often directed to some objectives (e.g., counting the number of ball passing actions) that do not feature the criteria for ``should have been'' (e.g., spotting a person in an unusual costume).
Hence selective attention is in general cost-beneficial. 

\textbf{Gestalt grouping} is another mechanism of alphabet compression.
The human visual system intrinsically groups different visual signals into patterns that are usually associated with some concepts, such as shapes, objects, phenomena, and events.
This enables humans to remember and think about what is being observed with a smaller alphabet.
Although the possible number of letters in the output alphabet after gestalt grouping is usually not a small number, it is a drop in the ocean when comparing with the number of letters in the input alphabet that represents all possible variations of visual signals.
Occasionally gestalt grouping may lead to visual illusions, where two or more processes for different types of gestalt grouping produce conflicting output alphabets.
In general, gestalt grouping functions correctly most of the time, second by second, day by day, and year by year.
In comparison, visual illusions are rare events, and pictures and animations of visual illusion were typically hand-crafted by experts.

Figure \ref{fig:DI-Psy1}(c) shows an instance of gestalt grouping.
Using a 1D version of the chess board illustration as the example input letter, the gestalt grouping process transforms the input to a pattern representation of repeated light-and-dark objects.
Such a representation will require less cognitive load to remember and process than the original pattern of eight objects.
However, when a person attempts to reconstruct the original pattern, for example, in order to answer the question ``do objects 2 and 5 have the same shade of grey?'' it is likely that a reconstruction error may occur.
Although visual illustration (potential distortion) is an inevitable consequence of gestalt grouping (alphabet compression), it does not happen frequently to most people.
Information-theoretically, this suggests that it must be cost-beneficial for humans to be equipped with the capability of gestalt grouping.

The human visual system consists of many components, each of which is a sub-system.
The human visual system is also part of our mind, which is a super-system.
In addition to the human visual system, the super-system of human mind comprises of many other systems featuring cost-beneficial data processing mechanisms.
Humans' \textbf{long-term memory} involves a process of representation for retention and another process of reconstruction for recall.
The former facilitates alphabet compression, while the latter introduces potential distortion.
The reason that most humans do not have photographic memory may likely be because the photographic mechanism would be less cost-beneficial.

\begin{figure}[t]
\centering
\includegraphics[width=\columnwidth]{Figures/DI-Psy}\\
(a) a generic data intelligence workflow in human mind\\[2mm]
\includegraphics[width=\columnwidth]{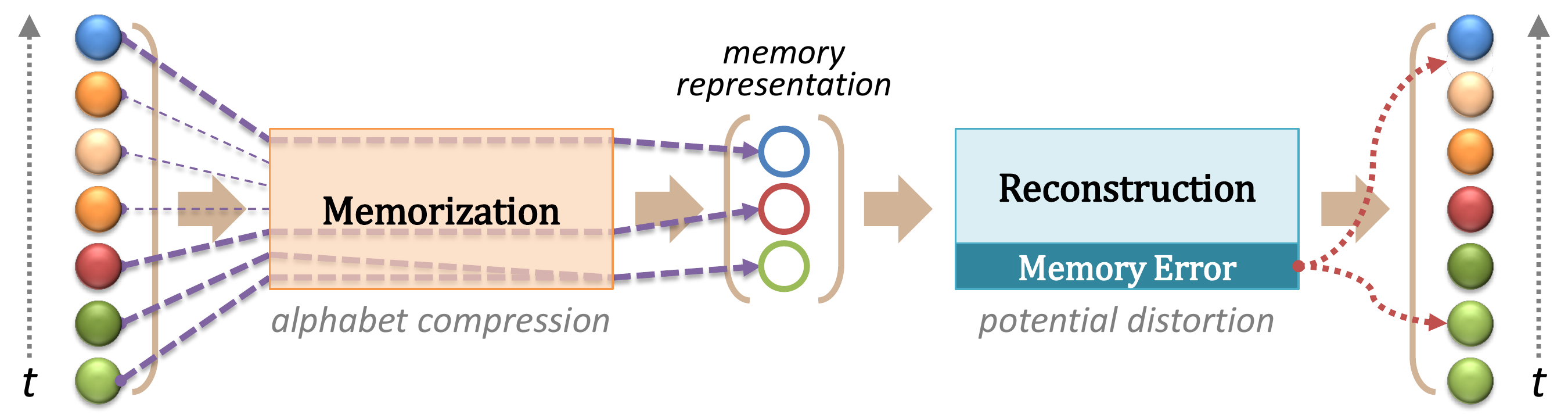}\\
(b) memory representation and memory errors\\[2mm]
\includegraphics[width=\columnwidth]{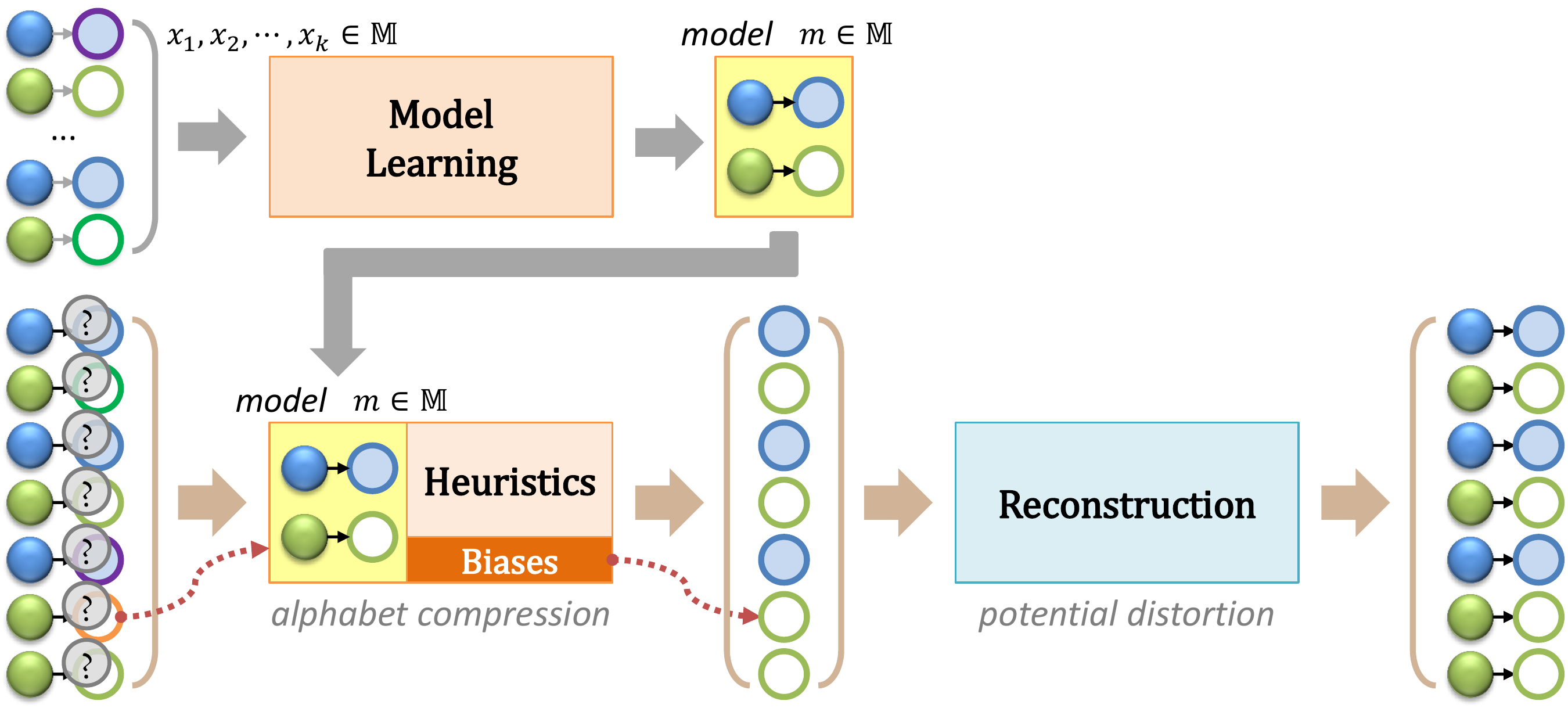}\\
(c) heuristics and biases
\caption{\emph{Memorization} and \emph{Heuristics} are two major phenomena in human cognition. Information-theoretically, they feature significant alphabet compression and are cost-beneficial, but may sometimes cause potential distortion in the forms of \emph{memory errors} and \emph{cognitive biases}.}
\label{fig:DI-Psy2}
\end{figure}

Figure \ref{fig:DI-Psy2}(b) shows an instance of memorization.
A sequence of temporal events may be captured by a person's mind, and an internal representation is created.
It is likely that this representation records only the main events and the important features of these events.
Hence a significant amount of alphabet compression enables humans to keep the cost of memorization very low.
During the memory recall, the sequence of events are reconstructed partly based on the recorded memory representation and partly based on the person's past experience and knowledge of similar sequences of events.
Some details may be added in inadvertently and some may be omitted, which leads to memory errors.

\textbf{Heuristics} are powerful functions for humans to make judgments and decisions intuitively and speedily.
If these functions are collectively grouped together into a system, each heuristic function can be considered as a sub-system.
Almost all decision-making processes facilitate alphabet compression since the output alphabet that encodes decision options usually has a smaller amount of uncertainty (i.e., entropy) than the input alphabet that encodes the possible variations of data, for which a decision is to be made.
Furthermore, although every situation that requires a judgment or decision mostly differs from another, somehow, humans do not maintain a decision sub-system for every individual situation.
This would not be cost-beneficial.
Instead, we develop each of our heuristic functions in a way that it can handle decisions in similar situations.
Grouping some functions for different situations into a single heuristic function is also a form of alphabet compression.
This is the consequence of learning. The potential distortion caused by such functional grouping is referred to as biases.
In a biased decision process, a relatively generic heuristic function produces a decision that is significantly different from an ideal function specifically designed for the situation concerned.

Figure \ref{fig:DI-Psy2}(c) shows instances of heuristics and biases.
Recall the model development workflow in Figure \ref{fig:DI-ML}.
It is not difficult to observe the similarity between the two workflows.
Figure \ref{fig:DI-Psy2}(c) depicts two major workflows, \emph{Model Learning} and \emph{Heuristics/Biases}.
A set of letters in $\mathbb{X}$ represent the past experience of some causal relations.
A model of the causal relations is learned gradually from the experience.
It is then used to make inference about the effects when some factors of causes are presented.
Because the heuristic model is intended to make decisions without incurring too much cognitive cost and may have been learned with sparse experience, the model may feature some overly simplified mappings from causes to effects.
Some of such over-simplification may be acceptable while others may not be acceptable.
The unacceptable over-simplifications are referred to as biases. 
The criteria for defining the acceptability usually depend on some contextual factors, such as purposes, tasks, consequences, social conventions, and so on. 

Today, many data intelligence systems are computer systems that employ functions, algorithms, and models developed manually or semi-automatically using machine learning.
Similar to human heuristics, their design, development, and deployment are intended to make decisions in a cost-beneficial way.
Also similar to human heuristics, they facilitate alphabet compression as well as introduce potential distortion inherently.
Biases may result from an over-generalization of some local observations to many situations in a global context, but they can also result from a simplistic application of some global statistics to individual situations in many local contexts \cite{Streeb:2018:book}.
Whilst there are cognitive scientists specialized to study human biases, we also need to pay more attention to computer biases.

A humans' heuristic model is learned typically over a long period and with controlled or uncontrolled exposure to training data and environments (e.g., learning in schools and judging traffic conditions).
Recall the discussions at the end of Section \ref{sec:Model}, with the rapid advances of virtual realty and virtual environment technology, we can design and create virtual training data and environments for developing special-purpose human models.

\section{Relating the Metric to Languages and News Media}\label{sec:Languages}
Consider all words in a language as an alphabet, and each word as a letter.
The current version of the Oxford English Dictionary consists of more than 600,000 words \cite{OED:2018:web}.
It is a fairly big alphabet.
Based on the maximal amount of entropy, they can be encoded using a 20-bit code.
The actual entropy is much lower due to the huge variations of word frequencies.

Nevertheless, in comparison with the variations of the objects, events, emotions, attributes, etc., this alphabet of words represents a massive alphabet compression.
For examples, using the word ``table'' is a many-to-one mapping, as the input alphabet contains all kinds of furniture tables, all kinds of row-column tables for organizing numbers, words, figures, etc., actions for having or postponing all kinds of items in a meeting agenda, and so on.
In fact, the Oxford English Dictionary has 28 different definitions for the word ``table'' \cite{OED:2018:web}.
When including the quotations, the word ``table'' is described using more than 34,000 words.
Interestingly, despite the ubiquity of alphabet compression, readers or listeners can perform, in most situations, the reverse mapping from a word to a reality at ease.
The potential distortion is not in any way at the level as suggested by the scale of one-to-many mappings.
Similarly to what has been discussed in the previous sections, it must be the readers or listeners' knowledge that changes the global probability distribution of the input alphabet to a local one in each situation.

We can easily extrapolate this analysis of alphabet compression and potential distortion to sentences, paragraphs, articles, and books.
With a significant amount of alphabet compression and a relatively small chance of potential distortion, written and spoken communications can enjoy relatively low cost by using many-to-one mappings.
This is no doubt one of the driving forces in shaping a language.

\begin{figure}[t]
\centering
\includegraphics[width=\columnwidth]{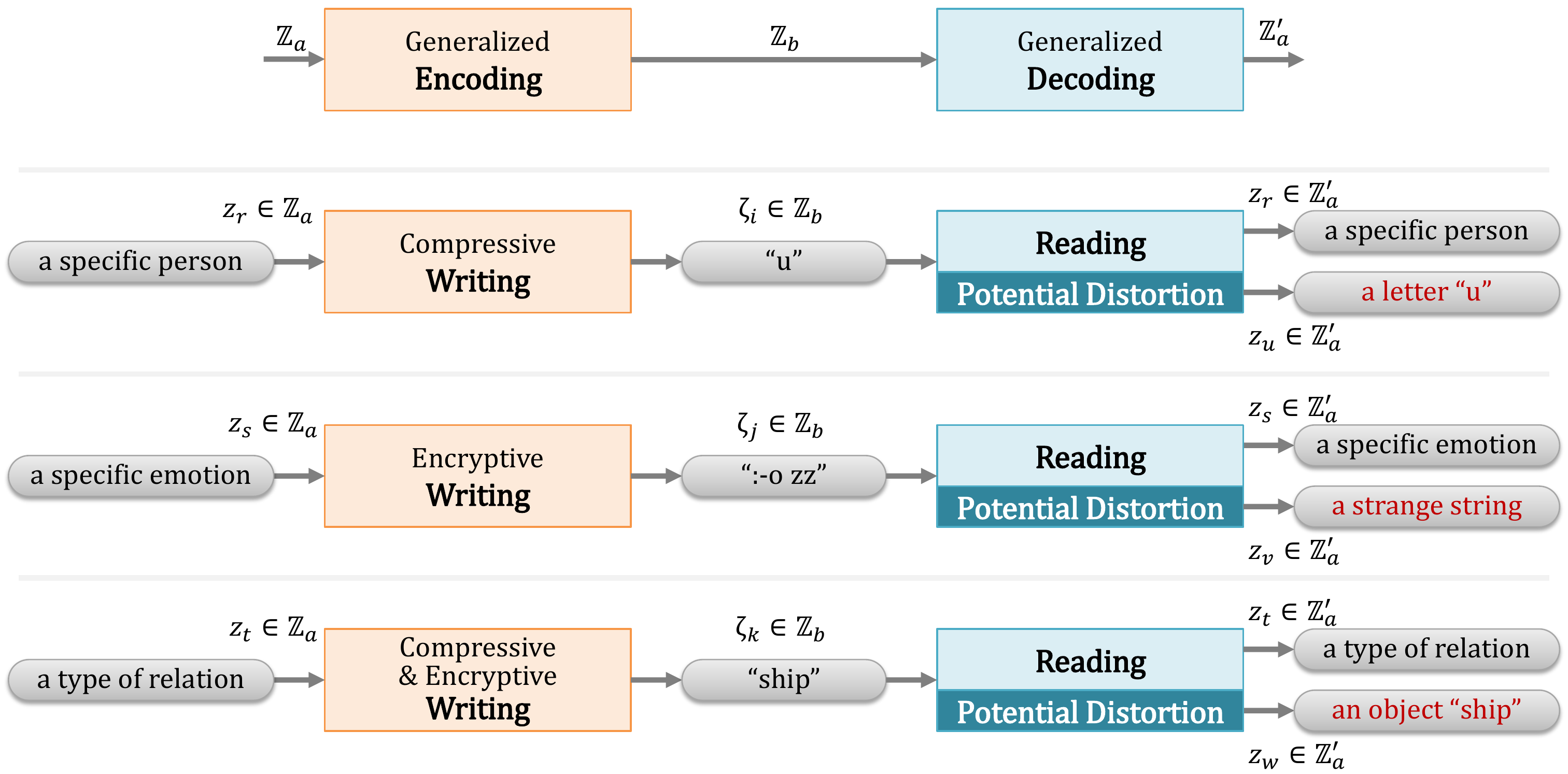}
\caption{Examples of compressive and encryptive encoding in  textual communications in some contemporary media such as texting.}
\label{fig:DI-Lan}
\end{figure}

Not only can written and spoken communications be seen as data intelligence workflows, they also feature processes exhibiting both encryption and compression at the same time, such as the creations and uses of metaphors, idioms, slangs, clich\'{e}s, puns, and so on.
To those who have the knowledge about these transformed or concealed uses of words, their decoding is usually easy.
These forms of figurative speeches often seem to be able to encode more information using shorter descriptions than what would be expressed without the transformation or concealment.
In addition, the sense of being able to decode them brings about a feeling of inclusion, enjoyment, or reward.
Figure \ref{fig:DI-Lan} shows three examples of ``words'' commonly encountered in textual communications through the media of mobile phone apps and online discussion forums.
They fall into the same generalized encoding and decoding workflow that was also shown in Figures \ref{fig:DI-Psy1}(a) and  \ref{fig:DI-Psy2}(a).
The cost-benefit analysis can thus  be used to reason about their creation and prevalence.

In many ways, these phenomena in languages are rather similar to some phenomena in data visualization.
A metro map typically does not convey the geographical locations, routes, and distances correctly.
All routes are deformed into straight lines with neatly-drawn corners for changing directions.
Such transformed and concealed visual encoding of the reality has been shown to be more effective and useful than a metro map with geographically-correct representations of locations and routes.
From an information-theoretic perspective, this is a form of cost-benefit optimization by delivering more alphabet compression and cost reduction using such encoding and by taking advantage of the fact that human knowledge can reduce potential distortion in decoding.

\begin{figure}[t]
\centering
\includegraphics[width=\columnwidth]{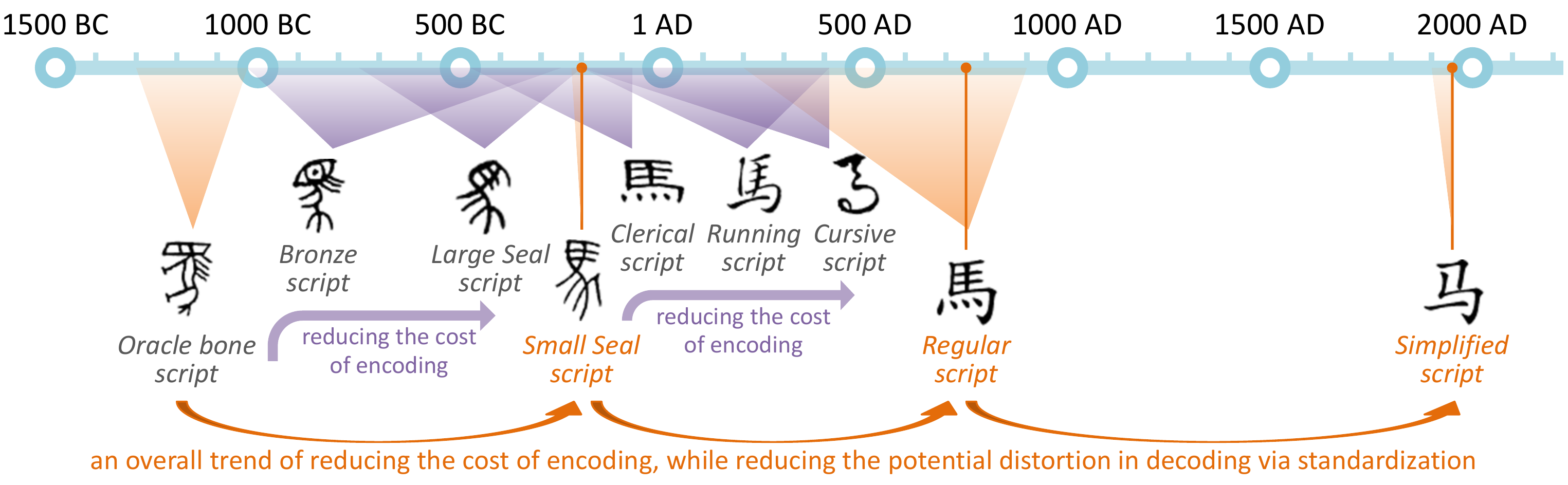}\\
(a) the historical evolution of of Chinese word horse (pronounced as ``ma'')  \\[2mm]
\includegraphics[width=\columnwidth]{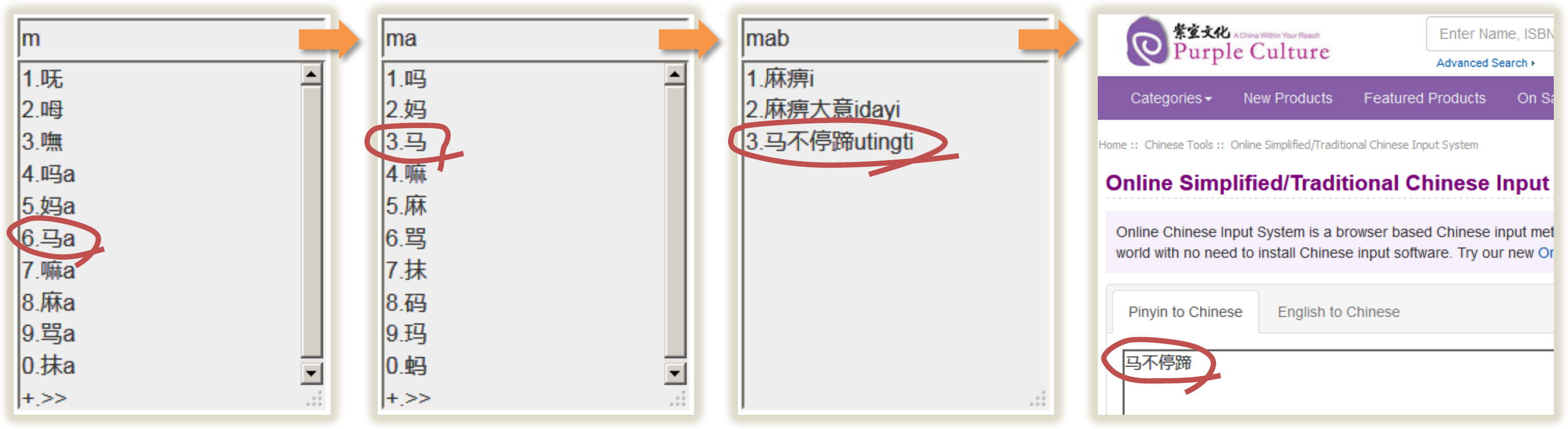}\\
(b) inputting a Chinese idiom starting with the character horse using a pinyin input tool \cite{PurpleCulture:2018:web}.
\caption{An example showing the reduction of the cost of writing Chinese characters over three millennia. It is common for modern input tools to make recommendations based on a user's partial inputs by using an entropic algorithm.}
\label{fig:Chinese}
\end{figure}

Languages have indeed been evolved with cost-benefit optimization in mind!
For example, Chinese writing, which uses a logographic system, has been evolved for more than three millennia \cite{Cao:2012:book,Hu:2014:book}.
As illustrated in Figure \ref{fig:Chinese}(a), the early scripts, which exhibit characteristics of both pictograms and logograms, bear more resemblance to the represented object than the late scripts.
Hence, the early scripts enable relatively easier reconstruction from logograms to the represented objects,  incurring less potential distortion in reading for readers at the era without systematic education.

The evolution of the Chinese writing system was no doubt influenced by many time-varying factors, such as the tools and media available for writing, the number of logograms in the alphabet, the typical length of writings, the systematization of education, standardization, and so on.
Information-theoretically, the evolution features a trend of decreasing cost for encoding and increasing potential distortion in decoding.
However, the improvement of education has alleviated the undesirable problem of increasing potential distortion.

The standardization of the seal scripts in the Qin dynasty (221 BC -- 206 BC) can be viewed as an effort for reducing the potential distortion and the cost in decoding.
The regular script, first appeared in 151 -- 230 AD, became a de facto standard in the Tang dynasty (618 -- 907 AD).
There are currently two Chinese writing systems, namely \emph{traditional} and \emph{simplified} systems, used by Chinese communities in different countries and regions.
The former, which was based on the regular script, is the official system in Taiwan, Hong Kong, and Macau, and commonly used in the Chinese Filipino community.
The latter, which was introduced in the 1950s, is the official system in mainland China and Singapore, and is commonly used in the Chinese Malaysian community.

Information-theoretically, the existence of the two writing systems is obviously not optimal in terms of cost-benefit analysis.
However, with the aid of modern computing and internet technologies, the cost of encoding and the potential distortion in dealing with two writing systems have been reduced significantly in comparison with the time without such technologies.
Most of modern digital writing systems have an entropic algorithm for character recommendation, and many support both traditional and simplified systems simultaneously.
Meanwhile many web sites in Chinese allow readers to switch between the two systems at the click of a button.
In many countries and regions where the Chinese language is extensively used, the co-existence of the two systems appears to be gaining ground, perpetuating the dissemination of information more cost-beneficially than the period when the modern technologies were not yet available and each country or region had to restrict itself to a single system.

As exemplified in Figure \ref{fig:Chinese}(b), the large proportion of encoding is now done on computers and mobile phones.
The encoding is truly a human-machine cooperation.
In Figure \ref{fig:Chinese}(b), a user needs to enter a Chinese idiom with four characters, which individually are translated as horse-not-stop-hoof.
It is a figurative speech for ``continuously making progress'' or ``continuously working hard''.
With the full pinyin spelling of each of the four characters, one online pinyin input tool \cite{ChineseTool:2018:web} showed 24 optional characters (in the simplified system) for ``ma'',  25  for``bu'', 21 for ``ting'', and 39  for ``ti''.
The total number of combinations available for the pinyin string ``ma bu ting ti'' is thus 491,400.
Another online tool \cite{PurpleCulture:2018:web}, which provides traditional and simplified Chinese characters simultaneously, showed 57 optional characters for ``ma'',  66 for``bu'', 60 for ``ting'', and 126 for ``ti''.
The total number of combinations available for the string ``ma bu ting ti'' is thus 28,440,720.
However, using an entropic algorithm, the optional characters and compounding phrases can be organized probabilistically based on their usage frequencies.
If the user needs to enter only the character horse, it can be done with ``m6'' or ``ma3''.
By the time the user entered ``mab'', there were only three options left.
The idiom, which is spelled with 10 pinyon letters, requires only 4 key strokes ``mab3''.
In comparison with the hand-writing system in Figure \ref{fig:Chinese}(a), the cost of encoding is significantly reduced.

In parallel with the phenomenon of cost saving in encoding Chinese characters digitally, we also see the rising popularity of emojis.
At the center of data science, the technology of  data visualization is pumping out new ``words'' (e.g., multivariate glyphs), ``sentences'' (e.g., charts), and new ``paragraphs'' (e.g., animations) everyday.
The modern communication media and computer technologies will continue to reduce the cost of visual encoding.
The alphabets of the lexical units, synataxical units, and semantic units of visual representations are growing in size and complexity rapidly.
The cost-benefit metric in Eq.\,(\ref{eq:CBR}) suggests that efforts such as standardization, systematization, and organized education can alleviate the potential distortion in decoding visual representations of data.
A language of data visualization may emerge in the future.
\\

The progressive development of languages brought about the paradigm of organized communication of news \cite{Stephens:1988:book}.
Early enterprises of news involved messengers and an organizational infrastructure for supporting them.
Messengers carried spoken or written words about news, commands, appeals, proclamations, and so on from one place to another.
The Olympic event \emph{Marathon} is attributed to the Greek messenger Pheidippides in 490 BC who ran 42 kilometers (about 26 miles) from the battlefield of Marathon to Athens to report the victory.
Ancient Persia is widely credited to have invented the postal system for relaying messages.
The ancient Mongol army had a communication system called ``Yam'', which was a huge network of supply points that provided messengers with food, shelter, and spare horses.
At that time, the messages transmitted were very compact, and collectively, in a much smaller number.
The cost for delivering the messages was overwhelmingly more than that for consuming the messages.

\begin{figure}[t]
\centering
\includegraphics[width=\columnwidth]{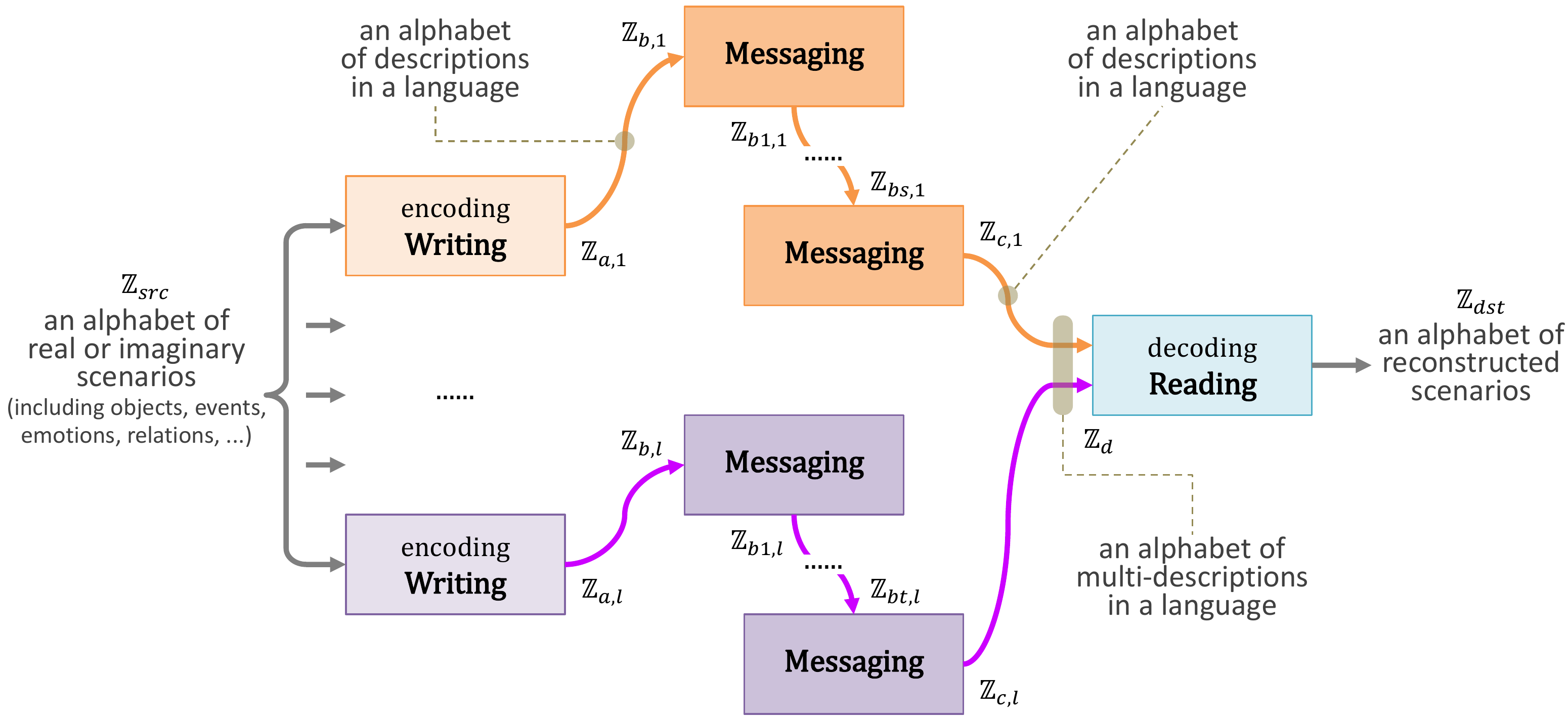}
\caption{A news communication workflow based on an infrastructure for relaying messages. Each message is relayed by a sequence of processes that are not expected to alter the semantics of the message, hence incurring no alphabet compression or potential distortion.}
\label{fig:Media-relay}
\end{figure}

\begin{figure}[t]
\centering
\includegraphics[width=\columnwidth]{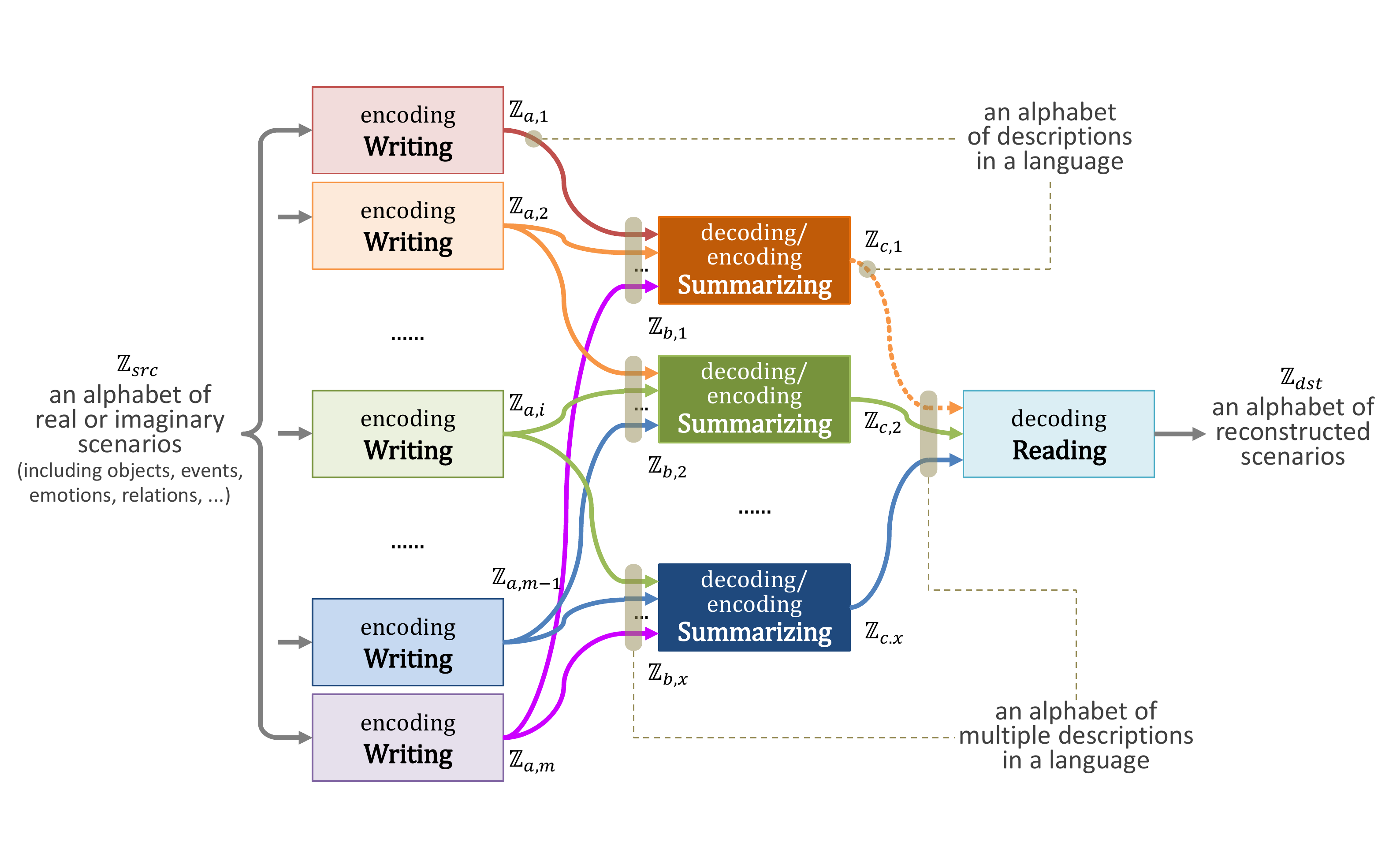}
\caption{A news communication workflow based on centralized media organizations, which deliver semantically-rich transformations from a large number of news reports to a selection of amalgamated and prioritized summary descriptions. The decoding and encoding in the intermediate processes result in both alphabet compression and distortion. While the workflow is commonly used for communicating with a large audience, as an example, only one reader is shown in the figure for the purpose of clarity.}
\label{fig:Media-centre}
\end{figure}

\begin{figure}[t]
\centering
\includegraphics[width=\columnwidth]{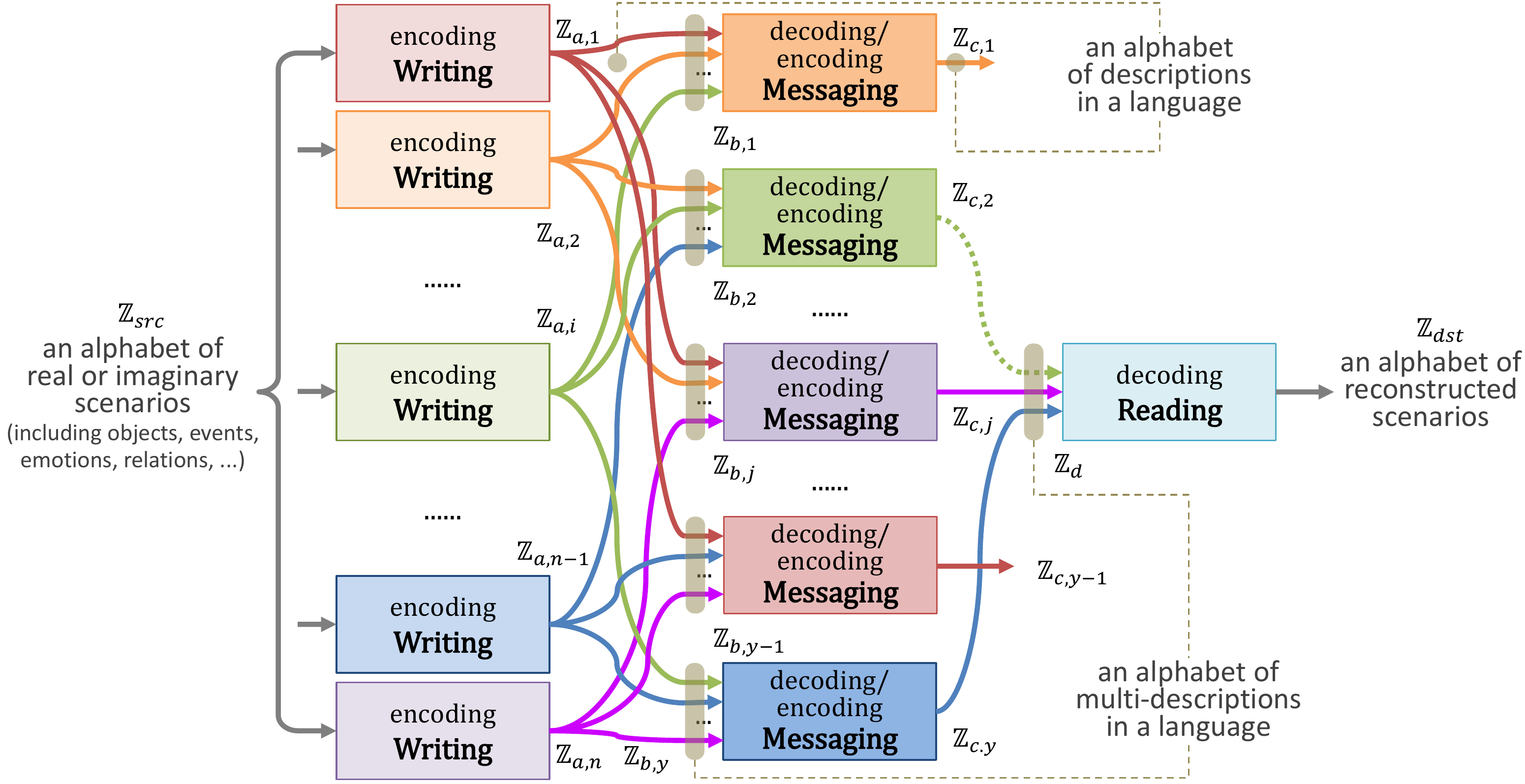}
\caption{A news communication workflow based on social media, where numerous new entities can perform semantically-rich transformations.}
\label{fig:Media-crowd}
\end{figure}

In contrast, the contemporary news media, which include newspapers, televisions, web-based news services, email newsletters, and social media, generate and transmit astonishingly more messages than the ancient news enterprises.
The cost of consuming all received or receivable messages is beyond the capacity of any organization or individual.
From an information-theoretic perspectives, we can consider the cost-benefit of three typical workflows of news communications as shown in Figures \ref{fig:Media-relay}, \ref{fig:Media-centre}, and \ref{fig:Media-crowd}.

Figure \ref{fig:Media-relay} illustrates a news communication workflow based on an infrastructure for relaying messages. 
This workflow encapsulates a broad range of mechanisms for delivering news across a few millennia, from ancient messengers on foot or horseback to postal mails, telegraphs, and telephones.
After the initial transformation from a real-world alphabet to a description alphabet of, for example, written messages (shown) or spoken messages (not shown), the intermediate transformations focus on transporting messages (i.e., letters in a description alphabet $\mathbb{Z}_{a,i}, i=1, 2, \ldots, l$).
In many situations, the intermediate transformations may involve additional encoding and decoding processes, for instance, digital-analogy conversions in telegraphs and telephones, and encryption and decryption for confidential messages.
Nevertheless, any semantic transformation is expected to take place only at the writing stage at the beginning and reading stage at the end of the workflow.
In other words, the intermediate transformations from $\mathbb{Z}_{a,i}$ to $\mathbb{Z}_{b[1],i}, \ldots, \mathbb{Z}_{b[\tau],i}$, and then to $\mathbb{Z}_{c,i}$, are designed to have a zero-amount of alphabet compression and a zero-amount of potential distortion.
Here $\tau>0$ is a path-dependent integer.
In terms of the cost-benefit metric for data intelligence, these intermediate relaying transformations bring about zero-amount of benefit information-theoretically.

Figure \ref{fig:Media-centre} illustrates another typical workflow for communicating news to a broader audience \cite{Klapper:1960:book}.
In this workflow, a number of centralized media organizations, such as newspapers, radio stations, and television stations, systematically gather a large collection of descriptions from many sources and produce summary bulletins to be consumed by many readers, listeners, and viewers (only one reader is shown in the figure).
The history of such broadcasting workflows can be traced back to ancient Rome around 131 BC, when \emph{Act Diurna} (Daily Acts) were carved on stone or metal and presented in public places; and the Han Dynasty of China (206 BC--220 AD), when central and local governments produced \emph{dibao} (official bulletins) and announced them to the public using posters and word of mouth.
Information-theoretically, the messages in the alphabets from differences sources, i.e., $\mathbb{Z}_{a,i}, i=1, 2, \ldots, m$, are amalgamated together to create more complex alphabets for multi-descriptions, $\mathbb{Z}_{b,j}, j=1, 2, \ldots, x$.
Each centralized organization performs semantically-rich transformations that feature a huge amount of alphabet compression in selecting, summarizing, annotating, and enriching the original messages in the input alphabet, resulting in an output alphabet of descriptions $\mathbb{Z}_{c,j}, j=1, 2, \ldots, x$.
To an individual reader (or a listener, or a viewer), there will be potential distortion when the person tries to imagine what was really said in $\mathbb{Z}_{a,i}, i=1, 2, \ldots, n$.

In comparison with the workflow in Figure \ref{fig:Media-relay}, the workflow in Figure \ref{fig:Media-centre} can handle more data at lower costs, especially from the perspective of cost-benefit per user.
Although the intermediate transformations may introduce potential distortion due to the many-to-one forward mappings, the amount of positive alphabet compression is not only necessary in most cases but also beneficial in general.
When the number of media outlets is easily countable, it is not difficult for a reader (or a listener, or a viewer) to become knowledgeable about the editorial coverage, styles, partiality, and so on.
Such knowledge may be used to make better reconstruction of the scenarios in $\mathbb{Z}_{src}$, e.g., by reading between the lines.
On the other hand, such knowledge may bias the decisions as to what to read (or to listen, or to view), leading to confirmatory biases.

Figure \ref{fig:Media-crowd} illustrates a contemporary workflow for communicating news through social media.
With the rapid development of the internet and many social media platforms, there are many more intermediate news entities that can perform semantically-rich transformations.
These entities can collectively access many more descriptions in the description alphabets, $\mathbb{Z}_{a,i}, i=1, 2, \ldots, n$.
Each of these entities can be as simple as a message relay process and as complex as a media organization.
They have a non-trivial amount of editorial power in selecting, summarizing, annotating, and enriching the original messages in the input alphabet.
They generate overwhelmingly more outputs, $\mathbb{Z}_{c,j}, j=1,2,\ldots,y$, than those centralized media organizations in Figure \ref{fig:Media-centre}.
For an individual reader (or a listener, or a viewer), it is no longer feasible to be aware of a reasonable portion of these $y$ outlets, and has to direct one's selective attention to a tiny portion of them.
In comparison with the workflow in Figure \ref{fig:Media-centre}, likely there is more selection but less summarization.

In summary, comparing the numbers of processes that can perform the initial transformations from $\mathbb{Z}_{src}$ in Figures \ref{fig:Media-relay}, \ref{fig:Media-centre}, and \ref{fig:Media-crowd}, we have $l \ll m \ll n$.
Comparing the numbers of intermediate media entities that can perform semantically-rich transformations in Figures \ref{fig:Media-centre} and \ref{fig:Media-crowd}, we have $x \ll y$.
Meanwhile, for each pathway (per user per message) from $\mathbb{Z}_{src}$ to $\mathbb{Z}_{dst}$, the cost of the workflow in Figure \ref{fig:Media-relay} is significantly higher than that in Figure \ref{fig:Media-centre}, which is further reduced massively by the workflow in Figure \ref{fig:Media-crowd}.
The main question will no doubt be the assessment of the \emph{Benefit} (i.e., \emph{Alphabet Compression} $-$ \emph{Potential Distortion}) in the cost-benefit metric.
The intermediate transformations for relaying messages (Figure \ref{fig:Media-relay}) incur little alphabet compression and potential distortion.
For both workflows in Figures \ref{fig:Media-centre} and \ref{fig:Media-crowd}, there is a trade-off between the huge amount of alphabet compression and the huge amount of potential distortion.
Some social scientists have suggested that the workflow based on social media may incur more potential distortion (e.g., uncertainty due to ``alternative truth'', confirmatory biases, and opinion polarization) \cite{Garrett:2009:JC,Brundidge:2010:JC,Gentzkow:2011:QJE,KnoblochWesterwick:2012:JC,Lee:2014:JC}.

Further qualitative and quantitative analysis will be necessary to estimate the \emph{Benefits} in bits. 
Nevertheless, as the cost-benefit metric suggests an optimization of the data intelligence workflows in news communications based on the fraction in Eq.\,(\ref{eq:CBR}), the cost reduction is certainly a significant factor in setting the trend.

\section{Conclusions}\label{sec:Conclusions}
In this chapter, we have applied the cost-benefit analysis, which was originally developed as an information-theoretic metric for data analysis and visualization workflows \cite{Chen:2016:TVCG}, to a range of data intelligence workflows in machine learning, perception and cognition, and languages and news media.
We have postulated and demonstrated the likelihood that the cost-benefit metric in Eq.\,(\ref{eq:CBR}) is a fundamental fitness function for optimizing data intelligence workflows that involve machine- and/or human-centric processes.
We have also made connections in abstraction between data intelligence workflows and data compression and encryption workflows.
Hence data intelligence can be seen as part of a generalized paradigm of encoding and decoding.
We can envisage the prospect that information theory can underpin a broad range of aspects in data science and computer science, and can be used to cogitate a wide range of phenomena in cognitive science, social science, and humanities.

While the broad interpretations that relate the cost-benefit metric in Eq.\,(\ref{eq:CBR}) to perception, cognition, languages, and news media will require further scholarly investigations to confirm, the reasonable possibility of its role as a fitness function in these broad interpretations as shown in this chapter provides further evidence that the metric can be used to optimize data intelligence workflows.
Of course, it may turn out that there are more general or applicable metrics with even broader explanatory power.
As long as this chapter helps stimulate the discovery of such new metrics, these broad interpretations are not just as useless as ``speculations'' that ``make interesting talks at cocktail parties'' \footnote{In his book \emph{A Short History of Nearly Everything} (p. 228, Black Swan, 2004), Bill Bryson told the story about a journal editor who dismissively commented on Candian geologist Lawrence Morley's proposition about the theory of continental drift.}.

\bibliographystyle{abbrv}
\bibliography{CBM}

\end{document}